%% file: main.tex
\documentclass[sigconf,10pt]{acmart}
\usepackage[english]{babel}
\usepackage{blindtext}
\renewcommand\footnotetextcopyrightpermission[1]{} 
\setcopyright{none}

\acmPrice{}

\usepackage{soul}
\usepackage{pifont}
\usepackage{xurl}
\usepackage{comment}
\usepackage{enumitem}

\usepackage[medium, compact]{titlesec}
\titlespacing*{\section}{1pt}{3.5pt}{2pt}
\titlespacing*{\subsection}{1pt}{3pt}{1.5pt}
\titlespacing*{\subsubsection}{1pt}{3pt}{1.5pt}

\input{new-commands}

\newcommand{\sys}{SFC\xspace}

\newcommand{\spfc}{SFC-P\xspace}
\newcommand{\onramp}{OnRamp\xspace}

\newcommand{\sqp}{BTS\xspace}
\newcommand{\sqps}{\sqp{}s\xspace}

\newcommand{\pct}[1]{$P{#1}$}
\newcommand{\wg}{{RPC}\xspace}
\newcommand{\wf}{{Hadoop}\xspace}
\newcommand{\ww}{{WebSearch}\xspace}

\titlespacing*{\section}{1pt}{3.5pt}{2pt}
\titlespacing*{\subsection}{1pt}{3pt}{1.5pt}
\titlespacing*{\subsubsection}{1pt}{3pt}{1.5pt}
\renewcommand{\paragraph}[1]{\noindent\textbf{#1}\ }

\usepackage{tikz}
\newcommand*\circled[1]{\tikz[baseline=(char.base)]{
            \node[shape=circle,draw,inner sep=0.5pt] (char) {#1};}}

\newcommand{\aditya}[1]{\todo[inline, color=brown]{aditya: #1}}

\newcommand{\jk}[1]{\todo[inline, color=cyan]{jk: #1}} 
\newcommand{\yiorgos}[1]{\todo[inline, color=violet]{Yiorgos: #1}}

\begin{document}
\title[\sys: Source Flow Control]{\sys: Near-Source Congestion Signaling and Flow Control}


\author{Yanfang Le$^1$,
Jeongkeun Lee$^1$,
Jeremias Blendin$^1$,
Jiayi Chen$^2$,
Georgios Nikolaidis$^1$,
Rong Pan$^3$,
Robert Soul\'{e}$^4$,
Aditya Akella$^{2,5}$,
Pedro Yebenes Segura$^1$,
Arjun Singhvi$^5$,
Yuliang Li$^5$,
Qingkai Meng$^6$,
Changhoon Kim$^7$, 
Serhat Arslan$^7$}
 \affiliation{%
   \institution{$^1$Intel\; $^2$UT Austin\; $^3$AMD\; $^4$Yale University\; $^5$Google\; $^6$Tsinghua University\; $^7$Stanford University}         
 }

\renewcommand{\shortauthors}{Y. Le et al.}

\input{sections/0abstract.tex}
\maketitle
\input{sections/1intro}

\input{sections/2motiv}  
\input{sections/3design}  
\input{sections/4impl}  
\input{sections/5eval}

\section{Conclusion}
This paper presents Back-To-Sender (BTS) and Source Flow Control
(SFC) as a novel use case of BTS. 
BTS provides the fastest possible signal of congestion from
switch ingress. SFC sends routable BTS packets from remote congestion points
to traffic sources and caches the information at near-source switches, 
creating a near-source sub-RTT signal-reaction loop while
avoiding HoL blocking in the switching fabric. SFC allows for
incremental, brownfield deployments via options like ToR-only deployments and
BTS-to-PFC conversion at the fabric edge. A
major standards body has begun the process of standardizing
SFC and its signaling mechanism.

\bibliographystyle{plain}
\bibliography{refer}

\appendix
\input{sections/appendix}
\end{document}

%% file: new-commands.tex
\usepackage{url}
\usepackage{graphicx}
\usepackage{latexsym}
\usepackage{amsmath}
\usepackage{bm}
 
\usepackage{amssymb}
\usepackage{amsfonts}
\usepackage{comment}
\usepackage{array}
\usepackage{xspace}
\usepackage{psfrag}
\usepackage[tight,small]{subfigure}
\usepackage{graphicx}
\usepackage{color}
\usepackage[linesnumbered,ruled,vlined]{algorithm2e}
\SetAlFnt{\small}
\SetStartEndCondition{ }{}{}%
\SetKwProg{Fn}{Procedure}{\string:}{}
\SetKwFunction{Range}{range}
\SetKw{KwTo}{in}\SetKwFor{For}{for}{\string:}{}%
\SetKwIF{If}{ElseIf}{Else}{if}{:}{elif}{else:}{}%
\SetKwFor{While}{while}{:}{fintq}%
\SetKwComment{Comment}{$\triangleright$ }{}
\AlgoDontDisplayBlockMarkers\SetAlgoNoEnd\SetAlgoNoLine%

\usepackage{enumitem}
\usepackage{alltt}
\usepackage{multirow}
\usepackage{array}
\usepackage{wasysym}
\usepackage{rotating}
\usepackage{datetime}
\usepackage{amsthm}
\usepackage{epsfig}
\usepackage{endnotes}
\usepackage{epstopdf}
\usepackage{booktabs}
\usepackage[]{todonotes} 
\usepackage{cleveref}
\usepackage{csquotes}
\usepackage[font=small,skip=0pt]{caption}
\usepackage{enumitem}
\setlist{nolistsep}

\newcommand{\ie}{\textit{i.e.}\xspace}
\newcommand{\eg}{\textit{e.g.}\xspace}

\definecolor{gray2}{rgb}{0.8,0.8,0.8}

%% file: sections/0abstract.tex
\begin{abstract}
State-of-the-art congestion control algorithms for data centers alone do not cope well with transient congestion and high traffic bursts. 
To help with these, we revisit the concept of direct \emph{backward} feedback from switches and propose Back-to-Sender (BTS) signaling to many concurrent incast senders. Combining it with our novel approach to in-network caching, we achieve near-source sub-RTT congestion signaling.
Source Flow Control (SFC) combines these two simple signaling mechanisms to instantly pause traffic sources, hence avoiding the head-of-line blocking problem of conventional hop-by-hop flow control.
Our prototype system and scale simulations demonstrate that near-source signaling can significantly reduce the message completion time of various workloads in the presence of incast, complementing existing congestion control algorithms. 
Our results show that SFC can reduce the $99^{th}$-percentile flow completion times by $1.2-6\times$ and the peak switch buffer usage by $2-3\times$ compared to the recent incast solutions.
\end{abstract}

%% file: sections/1intro.tex
\section{Introduction}~\label{sec:introduction}
Modern end-to-end congestion control schemes use rich
network state information, such as {\em In-Network Telemetry} (INT)~\cite{hpcc, powertcp} or detailed
delay dissections~\cite{swift, onramp}, instead of traditional
one-bit ECN~\cite{dcqcn, dctcp} or packet drops~\cite{reno_vegas} as congestion signals.
However, even with INT or detailed delay signals, these end-to-end approaches
experience a large round-trip time (RTT)-timescale signaling loop.
Congestion signals are carried by or derived from 
data packets that are themselves experiencing congestion or failure events along the forwarding path,
i.e., the signaling path itself is 
delayed by the on-going congestion. This can be disastrous. For instance, under incast, congestion queuing delay in a typical shared buffer switch can spike up to a millisecond, which is 1-2 orders of magnitude
larger than the congestion-free base RTT of data center networks (10-20$\mu$s)~\cite{swift, hpcc}. 

The coupling of congestion queuing and congestion signaling path
prevents the congestion control (CC) logic---either at the sender or receiver---from learning about the existence and precise degree of congestion in a timely manner.
As such, CC logic is forced to "smooth out" stale congestion signals and react conservatively so as to not induce unfairness in bottleneck sharing, congestion oscillation, or under-utilization of the bottleneck bandwidth.  
Ultimately, even modern CC schemes~\cite{hpcc, swift, powertcp} require multiple 
RTTs to fully detect and react to network congestion with each RTT being inflated by the large congestion queuing. 

A more desirable approach is to decouple the congestion signaling loop from the congested path. 
This helps signaling to be independent of the 
variable congestion delay that is the congestion control target. 
In this paper, we show how to achieve this decoupling, in particular, showing how to bound the signaling delay to be sub-base-RTT. 
We further show the implications of faster and precise feedback by designing a new and effective flow control approach.

Faster congestion signaling becomes more critical as Ethernet link speeds increase: 100GbE and 200GbE are being widely deployed in cloud DCs and 400GbE is common in AI training systems~\cite{amazonefa,googlecloudbandwidth}. 
The rapidly growing bandwidth-delay product (BDP) (a $400$GbE network with an RTT of $10 \mu$s has a BDP of $500$KB) means most message sizes fall within one BDP worth of bytes~\cite{bfc}. 
This means that at these link speeds, end-to-end CC is often too slow, i.e., it takes multiple RTTs, to react to congestion efficiently. 

Drawing inspiration from backward congestion signaling, notably ICMP source quench~\cite{icmp} and
IEEE QCN~\cite{qcn}, we introduce \textbf{BTS} {\em(Back-To-Sender)} as a simple approach to
sub-RTT signaling of network congestion. In contrast to
RTT~\cite{swift, timely}, ECN~\cite{dctcp}, or INT~\cite{hpcc} signals,
BTS is generated at switch ingress {\em prior to} data packets being enqueued at a congestion location, meaning that the BTS feedback loop delay is bounded by the
base network RTT, much smaller than the congestion-inflated sender-receiver RTT.  
Information from BTS signals could then be used to take a variety of precise actions CC~\cite{bolt}, flow control, fast-failover, etc.

We present one such example use of BTS, a flow control scheme called {\em Source Flow Control} \textbf{(\sys)}. 
\sys pauses the source of each flow as a reaction to the BTS signal. 
A congested \sys switch sends routable BTS packets with precise pause duration directly back to traffic sources. 
\sys thus avoids head-of-line (HoL) blocking in the switching fabric, the major limitation of hop-by-hop flow control today~\cite{irn,dcqcn}.
\sys achieves this using very few resources and without complex flow tracking logic in switches. QCN also uses back-to-sender signals from switches; but as a L2 CC, it adjusts the sending rate over multi-RTT AIMD (Additive Increase, Multiplicative Decrease) steps. In contrast, \sys immediately pauses all incast senders within sub-base-RTT, quickly reacting to incasts by modern line-rate transports.

To further reduce signaling delay, \sys \textbf{caches} the remote congestion information carried by BTS packets
at the sender-side Top-of-Rack (ToR) switches, which then uses the cached information to
instantly pause control new flows heading to the cached congestion points.
The BTS packets triggered by early incast senders opportunistically
propagate the congestion information to network ingresses and suppress later-coming incast flows close to their sources.
This sender-side caching effectively shrinks the signaling loop further down to server-ToR
one-hop RTT for most of incast senders. 

We have implemented the switch function of \sys targeting RoCEv2
networks 
on the Intel Tofino~2 programmable ASIC~\cite{tofino2}. 
A faithful implementation of the end-host functionality required by \sys needs small changes to  existing NIC designs or to end-host networking stacks. 
We show how \sys could work with today's NICs via an approximate realization of end-host functions that leverages features of existing RDMA NICs.

This paper aims to advance the state-of-the-art with the following contributions:
\begin{itemize}[leftmargin=*]
    \item BTS + caching: move congestion signaling near the traffic source.
    \item SFC: novel use case of BTS for low delay flow control.
    \item Concrete yet simple system design and implementation, with incremental deployment options (\S\ref{sec:implementation}).
    \item Simulations (\S\ref{sec:evaluation}) and theoretical analysis (\S\ref{app:theory}) showing the key aspects of BTS and SFC: (1) reducing switch buffer usage up to $3\times$ while avoiding throughput degradation, (2) hence improving application tail latency up to $6\times$, (3) atop modern congestion control (HPCC), flow control (OnRamp) and loss recovery (IRN) mechanisms. 
\end{itemize}

\noindent This work does not raise any ethical issues.

%% file: sections/2motiv.tex
\section{Novelty over Prior Work}\label{sec:motivation}

\sys leverages low-latency BTS pause signaling with in-network caching for direct flow control of traffic sources.
While \sys is the first approach combining the three building blocks to our best knowledge, BTS signaling draws on the long history of congestion control and flow control mechanisms. 
Below, we provide additional details that contextualize
the design of BTS and SFC.

\subsection{Limitations of Congestion Control}
End-to-end congestion control (CC) schemes today can reduce the impact of
sudden and transient congestion by using either conservative
slow start (TCP, iWARP~\cite{iwarp}) or RTS/CTS-style
solicitations (1RMA~\cite{akella20}, MPI~\cite{walker92}). 
Slow start under-utilizes available bandwidth, hurting FCT of small messages. 
Solicitation-based approaches pay an additional RTT cost per transfer.
To avoid the problems of slow start, many CC schemes, such as HPCC~\cite{hpcc}, DCQCN~\cite{ndp}, NDP~\cite{ndp}, or Homa~\cite{homa}, start transmissions at full line rate or set the initial window large enough to immediately fill up the bandwidth delay product
(BDP). 
This approach ensures minimal flow completion time when there is no congestion. 
However, upon incast, many concurrent line-rate senders cause high queue build-up and, eventually, packet loss.

The reaction of CC to line rate incast is slow. To make the total arrival rate \emph{equal to} the drain rate (e.g., line rate), the conventional heuristic (``cut rate or cwnd by half'' every RTT) needs $log_2$\emph{(\# of incast flows)} RTTs, easily exceeding 100us. Once the total rate matches the drain rate, the senders need to further decrease their sending rates close to \emph{zero} to drain the queue and release the shared buffer for other traffic sharing the switch (not just the queue). With CC, additional RTTs are required to control the total rate close to zero. 

Then, when contending flows stop and link bandwidth becomes available, CC needs multiple RTTs to ramp-up the rate. Even with a precise congestion signal like INT, the signal is delayed and stale proportionally to the congestion queuing delay; correspondingly, the CC logic is forced to conservatively react to the variably-delayed congestion signal to avoid congestion oscillation, needing more RTTs (often 10-100x of base RTT) to converge.

Flow control, especially paired with sub-base-RTT signaling, gives instant pause and resume capability, freeing up CC from the burden of multi-RTT decrements and increments upon transient congestion. While it’s possible to design a new CC scheme that reacts to BTS-signaled incast or transient congestion, in this paper, we take a simpler flow control approach similar to like OnRamp~\cite{onramp}, which tackles incast and transient congestion via flow control while CC handles the equilibrium behavior of long flows.

\subsection{Limitations of Hop-by-Hop Flow Control}
\label{sub:bts-limit}
Due to its simplicity and fast reaction, there has been increased
interest in using flow control (FC) to either augment end-to-end CC (e.g., PFC~\cite{pfc}, OnRamp~\cite{onramp})
or fully replace CC (e.g., BFC~\cite{bfc}).

PFC is a widely used hop-by-hop FC for lossless RoCEv2~\cite{rocev2}.
PFC pauses the intermediate link neighbor of a congested switch port/queue.
As documented in~\cite{pfc-problems}, one main issue is
head-of-line blocking due to the limited number of PFC priorities
(i.e., hardware queues).
Under high network load and incast, the PFC
back-pressure spreads congestion throughout the network core
and sometimes down to source ToR switches, effectively \emph{slowing down the
  entire network fabric}.

Backpressure Flow Control (BFC)~\cite{bfc}
addresses the HoL blocking issue by effectively pausing/resuming \emph{individual application flows} at every NIC or switch hop. To avoid flow collisions in a queue, BFC keeps track of active flows and unused queues in the switch ingress and dynamically map flows to the limited number of queues. Since per-flow occupancy of a queue is not known at ingress, BFC signals (ideally every) data packet dequeue events from egress to ingress via mirroring and recirculation. 
In our BFC implementation, we find that this operation 
consumes significant recirculation bandwidth leading to packet drops.
Mitigating this requires substantial changes in 
switching ASICs, such as a bookkeeping data structure implemented in the (otherwise fixed-function) queuing system. 
Furthermore, the number of hardware queues in switches cannot keep increasing proportionally with network scale (\#nodes, \#flows) in the foreseeable future. 
This may create a situation where there are more per-switch active flows (during bursts) than available queues, leading to flow collisions that degrade performance.

BFC aims to fully replace end-to-end congestion control, for which it needs
the entire network (switches and NICs) to be upgraded. In
contrast, \sys pauses traffic at the source directly and allows for
brown field deployments (e.g., initially only on ToR switches).

\subsection{Flow Control at Source}
\label{ssec:flow-control-at-source}
To avoid HoL blocking and complex flow state tracking in the switches, \sys simply sends the congestion data from the switches back to the senders, who naturally maintain per-flow state and flow-control individual flows.

OnRamp~\cite{onramp} took a similar approach of pausing individual flows at the source while using
One-Way Delay (OWD) measured at the receiver to derive the pause time duration. It aims to augment congestion control by reacting to incast and transient congestion faster than CC.
However, it still uses a sender-receiver end-to-end
signaling loop that is coupled with on-path congestion. While more
precise than RTT measurements, the one-way delay signal of OnRamp
reflects the past congestion experienced by packets at network egress,
compared to the current queuing time used in BTS. 

In fact, there is a range of possibilities for how the flow control signal is computed and how it is communicated to the sender, with OnRamp being one point in this space. To explore the trade-offs, below we compare OnRamp and three variants of BTS using micro benchmarks. At a high level, \emph{all four convey pause time duration metric for senders to perform flow control but differ in how to compute the metric and where the metric is computed and signaled from.}

{\bf (1) OnRamp} measures OWD at the receiver and computes the pause time duration for end-to-end flow control. We run the full simulation code from \onramp authors, including its smoothing algorithm that avoids oscillation.
{\bf (2) Egress BTS} signal is generated post-queuing at the congested switch egress, then sent back to the senders.
It carries the expected draining time $(\mathit{current\_q\_depth} - \mathit{target\_q\_depth})/\mathit{port\_speed}$ as the pause duration. 
Since this is generated at egress, the signal generation rate is capped by the PPS (packet per second) throughput of the congested link, approximately $port\_speed/MTU$. \onramp shares the same signaling PPS limitation. 
{\bf (3) Ingress BTS} signal is triggered before the data packet enqueues at the congested switch queue, while carrying the same metric -- the expected draining time -- as the pause duration. Since this is triggered at switch ingress, many concurrently arriving senders are immediately signaled back without being affected by the congestion queuing.
{\bf (4) Ingress BTS + Cache} is the essence of SFC. It takes the fastest possible signal (Ingress BTS) from the congestion origin and opportunistically caches the information at the network ingress (sender-side ToRs) to suppress later-coming incast senders close to them.

\begin{figure}[!t]
\begin{minipage}{1.0\linewidth}
    \centering
    \includegraphics[width=\textwidth]{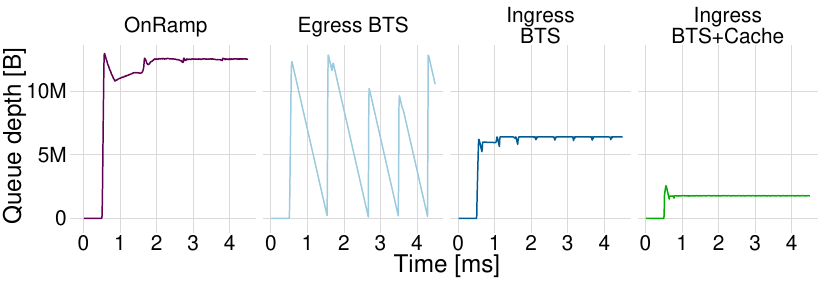}
    \caption{\small Effect of different signaling + flow control methods without congestion control. 63-to-1 incast senders starting within a 50us time window. A dumbbell topology with unlimited switch buffer is used with 10us NIC-to-NIC RTT.}
    \label{fig:sfc-bts-ingress-vs-egress-32flows-no-cc}
\end{minipage} 
\end{figure}

\subsubsection{FC handling of incast, without CC.}

We simulate benchmark experiments inspired by the \onramp paper~\cite{onramp}. We use a 2-switch dumbbell full-bisection topology to create a 63-to-1 incast where the senders and the receiver are connected to different switches via 100G links of 1$\mu$s delay.
One server starts sending at time $0$ms, the others start one by one spreading over a $50us$ synchronization window. 

In Fig.~\ref{fig:sfc-bts-ingress-vs-egress-32flows-no-cc}, we plot the queue depth change over the first $4$ms of the experiment to clearly see the fundamental effect of different flow control signaling schemes on queue depth, which directly impacts application tail latency. 
We assume infinite buffer size to simplify the analysis without worrying about the impact of drops and retransmissions on application performance. 
We first compare the FC signaling schemes \emph{without} end-to-end CC. Note that e2e CC is slower than FC, and cannot impact the peak buffer usage upon incast while it helps to reduce and converge the queue depth over time. We later enable CC to understand the combined behavior. 

The OnRamp graph in Fig.~\ref{fig:sfc-bts-ingress-vs-egress-32flows-no-cc} confirms the analysis in the \onramp paper: \enquote{a simple (straw-man) version of the \onramp algorithm that is intuitive, but has queue oscillations and the possibility of under-utilization in the presence of feedback delay}~\cite{onramp}. Hence it employs a smoothing mechanism that tunes the gain factor when there are more flows in the system, which takes an effect and stabilizes the queue depth within $2$~ms from the start. The equilibrium queue depth is $\approx13MB$. Note that CC is not enabled yet.

Egress BTS in Fig.~\ref{fig:sfc-bts-ingress-vs-egress-32flows-no-cc} doesn't implement any smoothing algorithm, hence it keeps oscillating as the straw-man version of \onramp does. The peak queue depth of Egress BTS is equal to the equilibrium state of \onramp, which shows that the effect of Egress BTS signal is analogous to that of receiver-side OWD. 
It makes sense because the signaling loop of Egress BTS is marginally shorter (just by 2x ToR-receiver 1hop link delay) than the e2e signaling loop of \onramp.

In contrast, Ingress BTS stabilizes itself quickly within $1$ms without the need for additional smoothing mechanisms. 
Egress BTS (as well as \onramp) needs at least $N$ dequeued packets at switch egress to inform $N$ different incast senders, while Ingress BTS doesn't suffer from such a $O(N)$ serialization delay of congestion signals.
Simultaneously arriving $N$ packets from different senders can trigger `Ingress' BTS almost at the same time. Thus, \emph{Ingress BTS makes a constant, low-delay feedback loop between the congested switch and the senders, achieving lower and stable queue depth} without an additional smoothing mechanism or CC.

To move the signaling even closer to the source, we employ a simple opportunistic caching
at the sender-side ToRs that
remembers the pause end time for each active incast receiver and pauses locally-attached senders. (See the next section for the full design.)  Fig.~\ref{fig:sfc-bts-ingress-vs-egress-32flows-no-cc} shows that Ingress BTS + Cache further lowers the queue depth and the reduction over Ingress BTS is proportional to the delay between sender ToR and receiver ToR. These observations are also coherent with our theoretical analysis presented in Appendix~\ref{app:theory}. In conclusion, Ingress BTS + caching provides the fastest and most precise congestion signal for the sender to react, thus we choose it as the congestion signal of \sys.

\paragraph{Incast synchronization.} We note caching doesn't help when the incast starts are perfectly synchronized. Perfectly synchronized incast is impractical in real systems; our survey didn't find a tight answer on the level of incast synchronization from real workloads but found a millisecond time window used to measure micro bursts in production systems like Millisampler~\cite{zhangimc17,Ghabashnehimc22}.

In our incast simulations, we use 50us (5x of 10us base RTT) or 100us as the default synchronize window. 
We later sweep various synchronization windows (as small as zero, modeling perfect synchronization) in \S\ref{sec:evaluation}. For the case of perfectly synchronized high-degree incasts, we introduce a simple incast estimation technique at sender-side ToR switches as an optional optimization (\S\ref{sec:lpf}). 

\begin{figure}[!t]
\begin{minipage}{1.0\linewidth}
    \centering
    \includegraphics[width=\textwidth]{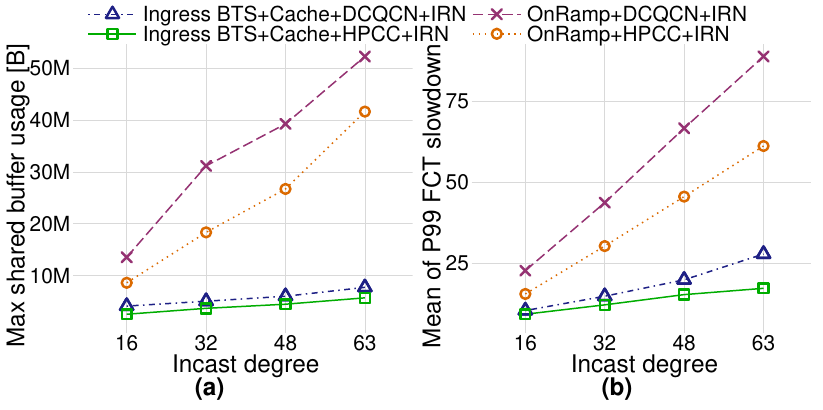}
    \caption{\small Effect of different FC + CC combinations, over various incast degrees.  Workload: 50\% network load (Hadoop traffic) + 8\% incast load with 50us incast synchronization. FCT is measured only for Hadoop flows.}
    \label{fig:dcqcn_with_hadoop}
\end{minipage} 
\end{figure}

\subsubsection{FC handling of incast, with CC.}

To see the augmented behavior of FC on top of CC, we plot the peak switch buffer usage of 2x2 combinations of FC and CC in Fig.~\ref{fig:dcqcn_with_hadoop}(a). For FC, we compare OnRamp and SFC (Ingress BTS + Cache) while running DCQCN (with congestion window) or HPCC as congestion controls (both implementations from~\cite{ns3-hpcc}).
We take 2~racks (each with 32 nodes) out of the 2-tier full-bisection topology used in the evaluation section~\ref{subsub:sim-perf} and generate an average 50\% network load of the cross-traffic using the message size distributions from \wf workloads~\cite{w4}. An 8\% worth of incast traffic load is added atop with varying incast degree from 16 to 63.
Again, unlimited switch buffer is assumed for this benchmark to clearly see the impact of FC + CC on buffer consumption, unaffected by RTO parameter settings.

Fig.~\ref{fig:dcqcn_with_hadoop}(a) shows the dominant impact of FC on switch buffer usage. Note we measure the total per-switch shared buffer usage that sums up the effect of the cross-traffic micro-bursts and the incast, unlike some other papers measuring per-queue depth. We find that while CC controls the queue depth down over time after each incast start (the timeline graph is not shown), the peak buffer usage is linearly proportional to the incast degree where the slope is governed by the FC in use. HPCC manages the cross-traffic's buffer usage better than DCQCN, but CC alone doesn't change the slope driven by incast. Thanks to its reduced signaling loop, Ingress BTS + Cache consumes much less buffer than \onramp over the entire range of incast degrees. 

In Fig.~\ref{fig:dcqcn_with_hadoop}(b), we measure the $99^{th}$ percentile of flow completion time (FCT) slowdown for each message size and then take the mean over all the message sizes. This tail latency metric is correlated with the peak buffer consumption, highlighting the benefit of Ingress BTS and SFC flow control on the application performance.

\subsection{Backward Switch Feedback}
\label{sec:backward-feedback}
Our work is inspired by prior proposals for backward notifications
from switches. These approaches were not  adopted widely in practice for various
reasons. First, ICMP Source Quench~\cite{icmp} was proposed for Internet
congestion control, but officially deprecated from the IETF because:
(1) it lacked a clear specification of how senders should react to the
quench, and (2) there was a lack of trust between devices in a WAN.
BECN~\cite{newman93,newman94} was another proposal; it is tied
to a specific congestion signal (ECN) and it does
not enable sub-RTT signaling since it requires the 1st RTT packets
to be ACK'ed by the receiver first to mark ECN on them.

QCN~\cite{alizadeh08} is similar to SFC in terms of pre-queuing congestion detection and backward notification. 
However, as an L2 "congestion control", it hasn't been well adopted in modern L3 data centers and overlaps with modern L4 congestion controls. 
QCN aims to determine the exact sending rate via multi-RTT AIMD, which is not fast enough to control switch buffer consumption pressured by high incast of modern line-rate HW transports.

Annulus~\cite{annulus} uses switch direct
feedback to mitigate long-RTT WAN traffic penalizing short-RTT
datacenter traffic.
For prototyping, Annulus hacked the L2 MAC learning switch feature to
generate switch backward signals and the paper nicely sets up the stage for BTS:
\emph{"(we) present a motivation for developing schemes that can report
INT signals directly to sender from switches"} for near-source
control loops. 

NDP~\cite{ndp} ensures constant low queueing upon incast by payload trimming and aggressive queue drop threshold selection (less than 1x BDP) but at the cost of many packet drops in the 1st RTT. Though the drops can be re-scheduled by the NDP receiver, the retransmissions can impact small message completion time and packet drops always complicate the transport design such as reorder buffer sizing and confusion with failure/black-hole drops. In addition, a burst of trimmed headers may turn the high BPS incast at the last-hop switch into a high PPS incast at the receiver NIC. PPS overload is one of the causes of NIC-side congestion and PFC triggering~\cite{jiang2020, collie}. In contrast, BTS distributes the congestion signal processing overhead to the senders.

Interestingly, NDP has Return-to-Sender (RTS, trimmed header to senders w/o any FC) as a preliminary
optimization to handle extremely large incasts, which may overflow the trimmed header queue (1x BDP) causing loss.
BTS is our answer to generic backward
signaling from switches as called out in Annulus and NDP. We further compare SFC and NDP quantitatively and qualitatively in Appendix~\S\ref{eval:ndp}.

%% file: sections/3design.tex

\section{\sys Design}
\label{sec:design}
\begin{figure}[!t]
    \centering
    \includegraphics[width=1.0\linewidth]{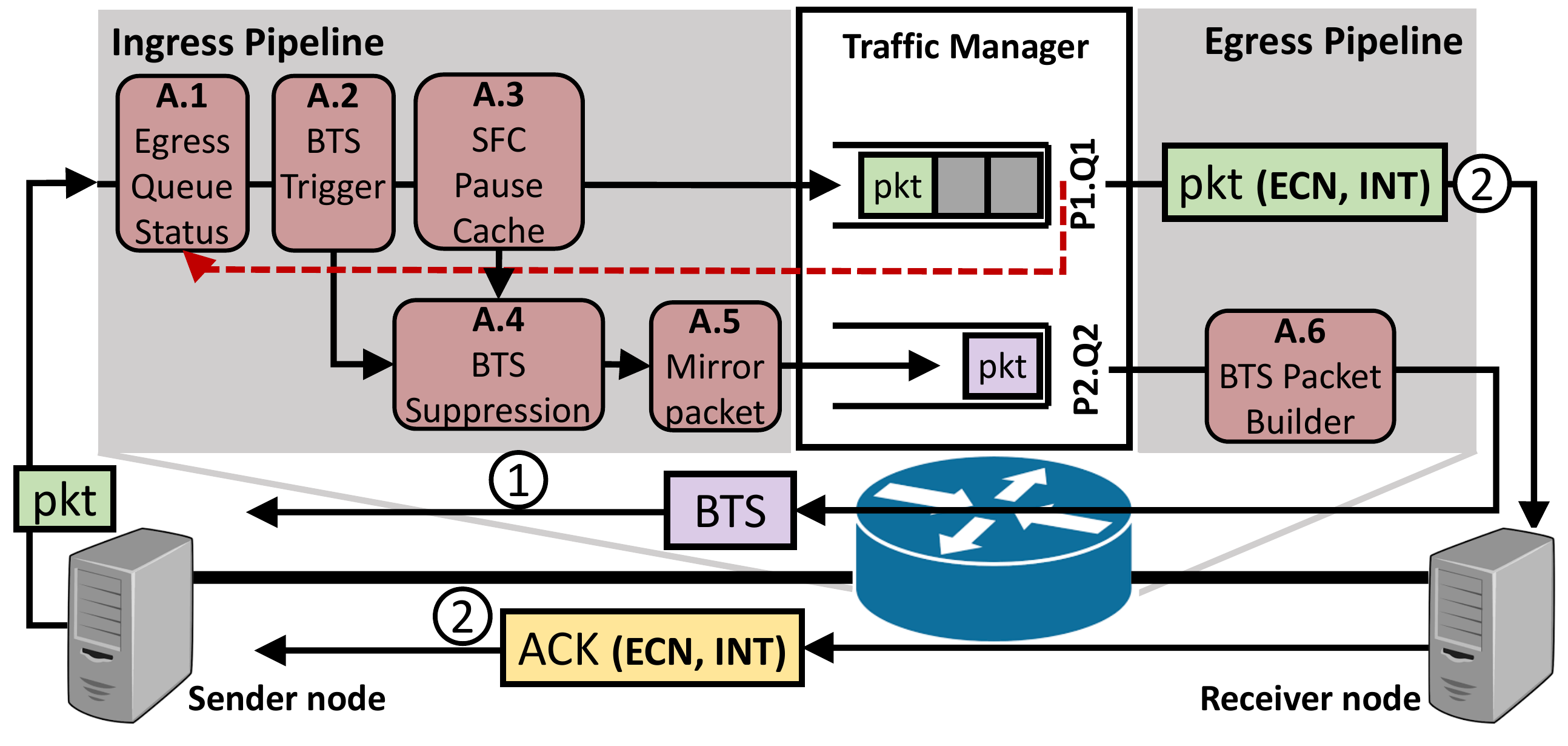}
	\caption{\small Switch pipeline of \sys. Upon congestion, a BTS is sent out (path \protect\circled{1}) before the data packet carrying ECN/INT reaches the receiver in the e2e signaling loop (path \protect\circled{2}). 
    }
    \label{fig:switchpipe}
\end{figure}

\sys is a flow control mechanism, where the switch ingress sends \sqp
with a pause duration to the sender of the traffic transiting through
a congested queue. The sender instantly
stops the affected traffic for a specified duration, moving the congestion queuing from
the switch buffer to the sender buffer. For near-source control, \sys caches the pause time carried by the BTS message at sender ToRs, which is highly effective when the incast senders are not perfectly synchronized. 
To handle tightly synchronized incasts, we also introduce an optional incast estimator that augments the caching mechanism. (\S\ref{subsec:cache}). 
Fig.~\ref{fig:switchpipe} depicts our switch pipeline.

\subsection{Switch Pipeline}~\label{subsec:pipeline}
Making queue information available 
at switch ingress (A.1 in Fig.~\ref{fig:switchpipe}) is key to \sys,
so that we can use it to
inform to all the senders of traffic being
forwarded to the congested queue about a congestion event, hence, enabling
highly parallelized sub-RTT signaling (\S\ref{subsec:ingress}).  

To further reduce signaling to one-hop RTT, we introduce a cache (A.3 in Fig.~\ref{fig:switchpipe}) 
at the sender-side ToR switch,
which stores the pause expiration time 
when it forwards a \sqp generated from a remote congested point.
Packets of later-coming senders at the same ToR look up the cached
information; if there is a match, the ToR switch sends \sqp (using
the cached pause time) back to the local senders,
even before their packets reach the remote congestion point.
Cache takes effect when the later incast senders start at least one RTT after the first incast sender at the same ToR.   
To suppress redundant back-to-back \sqp signals to the same source, we employ a simple Bloom filter (A.4)
with periodic clearing.

A data packet arriving at the switch ingress first looks up the forwarding
and QoS tables and retrieves its egress output port and queue (P1.Q1 in Fig.~\ref{fig:switchpipe}).
BTS is either triggered if the depth of the queue is over a threshold (A.2), 
or the SFC Cache indicates a remote congestion point towards its destination (A.3)
or (optionally) the incast estimator indicates a high traffic towards a remote congestion point (extension of A.3).
A \sqp is triggered back towards the packet sender if it is not suppressed by the Bloom filter (A.4).
A BTS packet is generated by mirroring the data packet, while trimming its payload (A.5). 
The header of the data packet is preserved in the mirror copy,
so that the receiver of BTS (i.e., the traffic sender)
can identify the source flow/connection to pause.

The mirror copy is turned into a L3-routable BTS packet at the
egress \sqp Packet Builder (A.6), where src/dst IPs are reversed
and a new UDP encapsulation header is added with its dst port set to
a predefined value.
The Packet Builder also computes the BTS pause time using the latest
information of the data packet's output queue (P1.Q1) available at egress (\S\ref{subsec:egress}). 
To avoid delay in the reverse path and ensure sub-RTT signaling,
BTS is routed through a high-priority queue similar to RoCEv2 CNP and ref.~\cite{timely, swift}.

Algorithm~\ref{alg:sfc-switch-cache} describes \sys in pseudo code;
note that pseudo code in \colorbox{gray2}{gray background} is specific to caching, which we
will cover later in \S\ref{subsec:cache}.

\begin{algorithm}[!t]
\caption{SFC switch logic \colorbox{gray2}{\strut w/ cache at sender-side ToR}}\label{alg:sfc-switch-cache}
\small
\colorbox{gray2}{$\mathit{cache[]} \gets$ {0}} \Comment*[f]{store pause end time}\\
\Fn{SfcIngress(pkt)}{
	\If{\colorbox{gray2}{isValid(pkt.bts) and pkt.bts.cacheable}}{
	\Comment{updates the cached pause end time by this BTS pkt.} 
	\colorbox{gray2}{$\mathit{key} \gets \mathit{[pkt.src\_ip]}$ }\\
	\colorbox{gray2}{$\mathit{cache[key] = max(cache[key], now + pkt.bts.pause\_time})$} \\
	\Comment{$now$ is switch local time} 
		\colorbox{gray2}{\textit{return}}
	}
	\Comment{here $pkt$ is a non-BTS data packet} 
        $isCongested \gets \mathit{queueDepth(pkt.queue\_id) > Q_{th} }$ \\
	\colorbox{gray2}{$\mathit{cache\_time} = \max(0, cache[pkt.dst\_ip] - now)$ } \\
	\If{isCongested \colorbox{gray2}{or cache\_time > 0}} { 
		$\mathit{suppr} \gets \mathit{bloomfilter(pkt.5tuple)}$ \\
		\If{$\mathit{not} \; \mathit{suppr}$ } {	
              \Comment{mirror $pkt$ with metadata ($mr\_meta$)} 
		    $\mathit{pkt.mr\_meta.queue\_id=pkt.queue\_id}$	 \\
		    \colorbox{gray2}{$\mathit{pkt.mr\_meta.pause\_time=cache\_time}$} \\
		}
	}
}
\Fn{SfcEgress(pkt)}{
	\If{pkt is a mirrored copy for BTS} {
 		$qd \gets \mathit{queueDepth(pkt.mr\_meta.queue\_id)}$ \\
		$\mathit{buildBtsHeader(pkt)}$ \Comment*[f]{reverse src/dst IP}\\

		$\mathit{pkt.bts.pause\_time} =$\colorbox{gray2}{$\max(pkt.mr\_meta.pause\_time,$} \\
$\;\;\;\;\;\;\;\;\;\;\;\;\;\;\;\;\;\;\;\;\;\;\;\;\;\;\;\;\;\;\;\;\;\;\;\;\;\;\;\;\;\;\;\;\;\;\;\;
	\mathit{(qd - Q_{Tg}) / port\_speed)}$ \\
		\colorbox{gray2}{$\mathit{pkt.bts.cacheable = isServerFacing(pkt.mr\_meta.queue\_id)}$} \\
	}
}
\end{algorithm}

\subsection{BTS Trigger at Ingress} \label{subsec:ingress}
\paragraph{Queue status at Ingress.}
Today, per-queue information is typically only available
at the egress pipeline of a switch, which, under congestion, packets (or a subset of packets) 
can reach only after experiencing queuing delay. 
However, \sys uses a direct queue-congestion feedback feature available
in recent switch designs, such as Intel Tofino~2~\cite{agrawal20}. 
This feature  
propagates depth information of certain queues, selected by the control plane, 
from the traffic manager to the ingress pipes\footnote{The propagation delay from the queue manager to the ingress table is less than the switch pipeline latency in our measurements.},
shown by the red dotted arrow in Fig.~\ref{fig:switchpipe}. 
Note that \sys at switch only stores a binary signal in a P4 register table~\cite{p4, p4spec} at ingress to indicate if the  queue depth is over the trigger threshold. The binary signal is read by data packets to make a BTS trigger decision (line 7 in Algorithm~\ref{alg:sfc-switch-cache}).

In triggering BTS, we use a simple static threshold mechanism (line 7) but
more sophisticated algorithms such as WRED, Proportional-Integral (PI)
controller~\cite{pan2013pie} and buffer-aware dynamic threshold are also possible in P4. 

\paragraph{BTS Suppression.}
\sys uses a suppression
mechanism based on a Bloom filter (line 10) to prevent a burst of \sqps to
be sent back to each of the incast sources.
The filter is reset periodically (e.g.,
every half RTT), to ensure that enough \sqps are generated for the
incast senders, keeping their pause times up to date. We tried reset period of half RTT, 1xRTT, 2xRTTs and we did not observe a big difference in terms of the application performance. We take the half RTT as the default reset period in the evaluation.

\subsection{\sqp Packet Construction at Egress}~\label{subsec:egress}
When a data packet is dequeued at egress, it carries the latest queue
depth data for its assigned queue, which is stored in a P4 register table and 
is then read by trimmed mirror copies to calculate the pause duration in BTS (line 16). 
The trimmed mirror copies are sent to a separate 
congestion-free egress port/queue (P2.Q2
in Fig.~\ref{fig:switchpipe}) and is then converted to a BTS packet (line 17).

\paragraph{Pause time calculation.}
The pause time $T_P$ is calculated as the time
needed to drain the congested
queue down to a target queue depth: $T_P = (qd - Q_{Tg})/R$,
where $qd$ is the current queue depth,
$Q_{Tg}$ is the target queue depth and $R$ is the port speed\footnote{We assume that this traffic class is
highest priority and can use most of the port speed
as in HPCC and Swift~\cite{hpcc, swift}.}  (lines 18-19, ignore the gray part).
The target queue depth
$Q_{Tg}$ is selected to ensure minimal queuing delay at full link
utilization.
The divide-by-$R$ operation can be implemented via a small memoization
table since a datacenter switch supports only few, well-known
link speeds. We express $T_P$ in microseconds
as we found that a finer-grained resolution does not
improve the system-level performance and is hard to implement in the NIC or host stack.

When the sender receives a \sqp, it pauses the 
targeted flow until the pause-end time, which is the sender's current time
$plus$ the pause duration specified in the \sqp.  If the sender
receives another \sqp for a connection that is currently paused,
its pause-end time is updated with the new one.

\paragraph{Parameter settings to prevent underrun.}
\sys uses two parameters, $Q_{Th}$ for triggering
a BTS packet and $Q_{Tg}$ as the target drain queue depth, where $Q_{Tg}<Q_{Th}$.
To achieve full link utilization, these two parameters must be sufficiently high while maintain the switch queue depth low.  

Let's set $Q_{Tg}=T_{F}*R$, where $T_{F}$ is the
feedback loop delay, i.e., the time it takes a packet to go from
the source to the congestion point plus the time it takes the BTS
frame to reach the source, and $R$ equals the port speed. 
Let's denote the gap between the two parameters as $X=Q_{Th}-Q_{Tg}$.

During congestion, the first BTS is triggered when the queue is built up to 
$Q_{Th} = T_F*R+X$. Let's 
set this moment as time~0 and assume the worst-case scenario (from the perspective of under-utilization) that no more
data packet is present at time~0, hence the BTS-triggering data packet is
the last data packet in the queue. 
The BTS packet carries a pause duration of $X/R$ and
reaches
the traffic source at time $T_F/2$, and by then
the queue depth at the congestion point
drains down to $T_F*R+X-(T_F/2)*R=T_F*R/2+X$. Since the pause duration is $X/R$, the data source can only start
sending new data at time $T_F/2+X/R$, at which point the queue depth
will be $T_F*R/2$. The new data reaches the congestion point
at time $T_F+X/R$, which is also the exact time at which the last data packet
(that triggered the previous BTS) is dequeued after spending $Q_{Th}/R$ time. 
Thus, setting $Q_{Tg}\geq T_{F}*R$ is recommended to avoid link under-utilization.
We choose 2xBDP as the triggering threshold and one BDP as the target queue depth in our experiments. 

\begin{figure}[!t]
\centering
\begin{minipage}[t]{1.0\linewidth}
	\centering
    \includegraphics{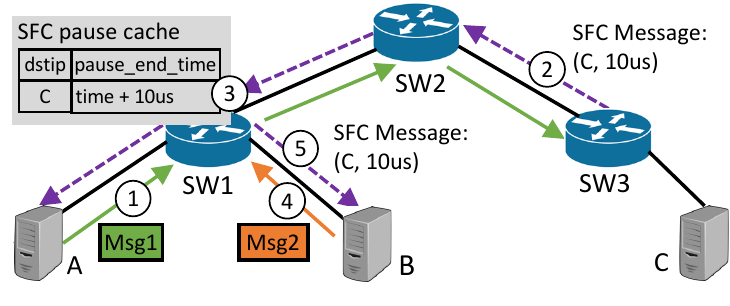}
	\caption{\small \sys Pause Cache. Host A sends traffic \protect\circled{1} that encounters
	congestion at SW3. SW3 sends back a BTS frame \protect\circled{2} which is cached
	at SW1 \protect\circled{3}. Host B then tries to send traffic to Host C \protect\circled{4}
	but receives a BTS frame triggered by SW1's cache \protect\circled{5}.
	}
    \label{fig:dstcache}
\end{minipage}    
\end{figure}

\subsection{Near-Source Caching and Estimation}
\label{subsec:cache}

The worst-case feedback loop delay for \sys is when the congestion point is
at the ToR switch port connected to the receiver host.
On the other hand, all traffic to a server is guaranteed to pass the
same congested ToR switch link in single-homing setups\footnote{We assume that the core network congestion can be handled by multi-path load balancing schemes ~\cite{plb, ndp}.}.
Based on this observation, we propose two techniques to move the congestion signal even closer to the senders: (1) opportunistically cache \sys pause information and (2) (optionally) monitor the traffic rate towards a destination server at the sender-side ToR switches.

\paragraph{Near-source caching.} 
The \sys Pause Cache block in the upstream (or just sender-side ToR) 
switches' ingress intercepts
\sqp packets (line 3, gray in Algorithm~\ref{alg:sfc-switch-cache}) and updates the cache with the IP
of the congested receiver (used as a congestion locator) and the pause end time (lines 4-5).
Since the src/dst IPs are swapped, the src IP of a BTS
packet carries the IP of the congested receiver. 
In Fig.~\ref{fig:dstcache}, switches
$SW1$ and  $SW2$ cache the information of a BTS packet
generated by $SW3$ when traffic from Host~A towards
Host~C encountered congestion. 
The $pause\_end\_time$ stored is the switch local
time advanced by the pause time carried in the BTS.

If a switch receives a data packet (e.g., Msg~2
in Fig.~\ref{fig:dstcache}) towards an IP in its cache and
the pause end time has not expired, it sends 
a \sqp to the data source, resulting in shorter
(just one hop) signaling delay.
\sqp carries a binary flag
indicating whether the congestion point is on a
server-facing port of a ToR switch (line 20).
This information is used by the upstream switches
to make a caching decision (line 3). 

Our theoretical analysis in Appendix~\ref{app:theory:congestion-caching} shows that SFC Pause Caching helps reduce congestion by $(1 - T_{ToR}/T_A)$ where $T_{ToR}$ is the propagation delay between the sender and the immediate ToR switch and $T_A$ is the propagation delay between the sender and the bottleneck.
Consistent with the theoretical results, our simulations in \S\ref{eval:simulation-robust} show that the cache is effective except for when the incast senders are perfectly synchronized. 

\paragraph{(Optional) incast estimator.} 
Caching cannot help when incast senders all start at the same time, e.g., less than the RTT between the senders and incast victim switch that is the initial BTS signaling delay. We present a simple extension to the caching mechanism that estimates incast by monitoring aggregate traffic towards a common receiver as seen by the sender-side ToR. 

There are multiple building blocks available in datacenter switches for this purpose, like meters or Discounting Rate Estimator (DRE)~\cite{conga}. We use the Low-Pass Filter (LPF) that implements DRE in Tofino and Tofino~2. LPF/DRE tracks the number of bytes seen in the last $\tau$ time, where the moving time window is approximated by exponential decaying. To get a rate, one can divide the byte estimation by $\tau$; but in our case, the byte number can be directly used to estimate the remote queue depth by simply subtracting $\tau * port\_speed$. Since each sender ToR doesn't have a global view in the 1st RTT (till BTS carries the global view from the congestion point), it may underestimate the remote queue depth and does not overestimate.

The estimated remote queue depth is used to compute the pause time (called \emph{local\_estimate}). Though omitted for brevity, a simple modification to Algorithm~\ref{alg:sfc-switch-cache} is sufficient, for example we take the max of cache\_time and \emph{local\_estimate} in line~13. Later our evaluation shows the effectiveness of this near-source estimation in avoiding drops upon a large, perfectly synchronized incast.
Unlike opportunistic caching that creates a table entry only when a BTS is received for a given destination, this local LPF estimation requires proactively created entries for potential incast destinations. 

We did not find evidence from production systems for such tightly synchronized incasts, so we expect this feature to be only occasionally needed and consider it to be optional.

\paragraph{}
\label{sec:lpf}

%% file: sections/4impl.tex
\section{System Implementation}
\label{sec:implementation}
We implemented \sys on an Intel Tofino~2~\cite{tofino2} switch
running SONiC switch OS~\cite{sonic}.
The standard SONiC features (e.g., L2/L3 forwarding, VRF, MAC learning) 
require more ingress resources than egress. 
To easily fit \sys with the standard features 
in the Tofino~2 pipeline, 
we implemented as many \sys functionalities (Fig.~\ref{fig:switchpipe})
in the egress pipe. This is another reason we placed the queue depth table,
pause time calculation, and \sqp packet builder
in the egress pipe.

\paragraph{\sqp forwarding.}
\sqp packet carries the 5-tuple of the original data packet but with
its source and destination IP pair swapped for reverse
forwarding to the data source. 
To pause RoCEv2 connections individually \sqp carries the Queue-Pair (QP) number which, together with the IPs, uniquely identifies the end-to-end connection of the original packet\footnote{RDMA packets don't carry a source QP,
hence we encode this in the source UDP port number.}. 

A naive approach to forward \sqp is sending it back to the 
incoming port of the original data packet, but a valid path
back to the sender is not guaranteed at the upstream switch of the data packet
during a route update and convergence period (caused by link failure or core switch upgrade).

Hence, we recirculate the BTS packets (constructed by egress BTS Builder) to the ingress where the forwarding tables decide a correct output port for the reversed IP.
This use of recirculation ports also allows the use of the latest egress-side queue state for computing the pause time.
The recirculation load is minimal because 1) BTS is triggered only for the queues congested over the triggering threshold, 2) redundant BTS packets are suppressed by bloom filter, and 3) payloads are trimmed.

\paragraph{Stateful Table Operations.}
The bloom filter used for BTS suppression requires periodic resets, performed by a timer-triggered packet generation engine or an optimized HW
function~\cite{agrawal20}. 
Our large-scale incast simulations show the worst-case Bloom filter
occupancy is under $150$ flows per switch~(\S\ref{eval:simulation-overhead}).
Its false positive rate is effectively zero in our implementation while taking up less than 0.1\% of available SRAM. 
Though rare, when false positives occur, impacted flows
may experience false suppression over multiple reset cycles. 
To prevent this, we add a version number (a simple counter) to the hash input that changes every reset cycle.

The SFC cache table matches on a key (dstIP, dscp) and stores a value (pause-end time). 
It needs a mechanism for adding new entries directly from the dataplane to handle
subsequent back-to-back BTS packet arrivals.  To that end, our Tofino
prototype and ns-3 simulations implemented a simple hash-addressed
register (32K entries per table), with low chance of hash collisions
for the level of table occupancy (under 15) observed in the
simulations. For better options of collision-free dataplane entry
insertion, Cheetah~\cite{barbette20} presented a P4 design that
maintains a stack of empty entry indices in the dataplane. 
Another option is the HW learning capability available in modern ASICs~\cite{agrawal20}.

\paragraph{\spfc: SFC-to-PFC conversion for today's NICs.}
\label{sec:design-star}
Since RoCEv2 transport is implemented in NIC hardware,
\sys requires a small but non-zero
hardware change to pause a QP as a reaction to \sqp without incurring SW processing delay.
We introduce \spfc (P for PFC),
which approximates the behavior
of \sys and is readily deployable with existing RDMA NICs. 
Today's RDMA NICs react to PFC and pause for a specified
amount of time per the IEEE 802.1Q standard, albeit on a
per-priority-queue rather than a per-flow basis. 
With \spfc, when a BTS packet from a remote congested switch reaches the ToR switch of the
sender, the sender-side ToR converts
the BTS packet to a PFC frame and sends this instead to the
sender NIC queue addressed by the BTS.
\spfc is used for system evaluations while we run \sys for simulations.

\paragraph{Implementability.} We note that the key building blocks of \sys -- payload trimming, mirroring, recirculation, telemetry, bloom filter, generic match-action (ACL) table, meter or rate estimator -- are
available in modern switches, whether fully programmable or not. Given the industry record of supporting QCN~\cite{qcn-allerton08, broadcom-qcn}, INT/HPCC~\cite{ifa} and trimming~\cite{trimming}, we believe \sys is readily implementable in commodity switches.

%% file: sections/5eval.tex
\section{Evaluation}~\label{sec:evaluation}
We evaluate \sys using a testbed and large-scale simulations using representative, public data sets on data center workloads to answer the following questions:
\begin{itemize}
    \item How does SFC manage switch buffer upon high incasts and avoid drops? (\S\ref{eval:testbed}, \S\ref{eval:simulation-perf})
    \item Does SFC avoid HoL blocking compared to PFC in real RoCEv2 systems? (\S\ref{eval:testbed})
    \item What is the improvement on application tail latency (FCT) compared to the state-of-the-art CC, FC and loss recovery mechanisms that are designed to handle large incasts? (\S\ref{eval:simulation-perf})
    \item Does \sys help also with micro bursts (no incast) or smaller incasts? (\S\ref{eval:simulation-perf})
    \item How robust is \sys over the system and simulation parameters? (\S\ref{eval:simulation-robust})
\end{itemize}

\subsection{Experimental Setup}
\label{subsec:setup}

\Cref{tab:eval-parameter} provides an overview on the key parameter combinations used in each environment.
\begin{table}[]
{\footnotesize
\begin{tabular}{p{3.15cm}|
				p{1.9cm}|
				p{2.3cm}}
  \textbf{Parameter} & 
  	\textbf{Testbed}~sec.~\ref{eval:testbed} & 
	\textbf{Simulation}~sec.~\ref{subsub:sim-perf}, appx. \ref{eval:simulation-overhead}, \ref{eval:cc-fairness}, \ref{eval:ndp} 
  	\\ \hline 
  Flow control schemes & 
  	\sys, PFC & 
  	\sys, \onramp 
  	\\ \hline
  Congestion control schemes & 
  	DCQCN+W & 
  	DCQCN+W, HPCC 
  	\\ \hline
  Packet loss handling & 
  	RoCEv2 GBN & 
  	IRN (selective retx) 
  	\\ \hline
  Workloads & 
  	incast, Hadoop & 
  	incast, RPC, Hadoop, WebSearch 
  	\\ \hline
  Topology & 
  	2-tier Clos, 1/2 switch, 48 nodes (100GbE), 3:2 over-subscr. & 
  	2-tier Clos, 8/16 switches, 512 nodes (100GbE), no over-subscr. 
  	\\ \hline
  Switch buffer & 
  	$16MB$, dynamic thresholding & 
  	$32MB$, dynamic thresholding
  	\\ \hline
  Base RTT & 
  	10$\mu s$ & 
  	10$\mu s$ (default), 16$\mu s$
  	\\
\end{tabular}
}
\caption{\normalsize Evaluation parameter overview}
\label{tab:eval-parameter}
\end{table}
We discuss the evaluation parameters in detail in the follow section. 

\paragraph{Flow control (FC) schemes.}
In simulation, we compare \sys against \onramp~\cite{onramp}.
For \onramp, we integrated the authors' implementation into our NS3 code and consulted them in finding good parameters for our topology. 
For \sys, we set the trigger threshold to $2\times$ BDP and the target queue depth to one BDP as described in Section~\ref{subsec:egress}.
As reference points, we also show results without flow control to indicate to which degree congestion control alone is able to control congestion in the network.

\paragraph{Underlying congestion control (CC) schemes.}
In simulations, we test with ECN-based DCQCN~\cite{dcqcn} and INT-based HPCC~\cite{hpcc}.
DCQCN is augmented with a static BDP-size window as introduced in~\cite{hpcc} (DCQCN+W).
We set the ECN threshold to one BDP and 
take the rest of the DCQCN+W parameters from the HPCC simulator~\cite{ns3-hpcc}.
For HPCC, we use the parameters (e.g., $\eta=0.95$) from the HPCC simulator~\cite{ns3-hpcc} since the network speeds are the same.

In the system evaluation in the testbed, we use DCQCN as underlying CC. 
The Intel E810 NIC implements DCQCN as a window-based CC for Write RDMA operations~\cite{intele810}, which we use in our workload generator.

\paragraph{Loss recovery mechanisms}
RoCE's Go-back-N~\cite{Guo:2016:ROC} (GBN) is the standard loss recovery mechanism in the industry, but is known for its bad performance in handling congestion drops. 
In the testbed experiments, we use GBN part of the NIC implementation.
IRN~\cite{irn} introduces selective retransmission for RoCE and is the state of the art in the literature. 
We use it as the loss handling mechanism in the simulations. 

\begin{figure*}[!htb]
\begin{minipage}{3.48in}
    \centering
    \includegraphics{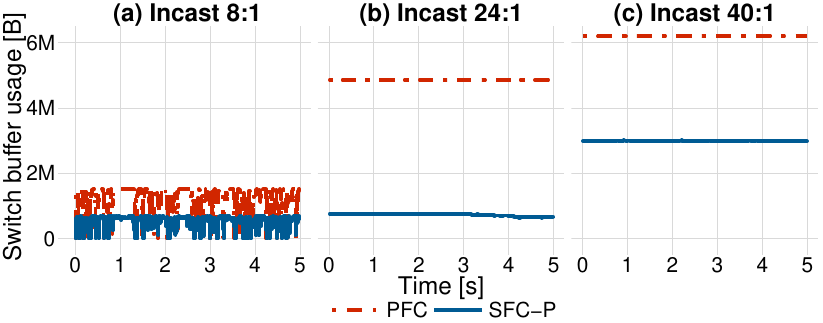}
    \captionsetup{margin={10pt,0pt}}
    \caption{Testbed: switch buffer usage by incast degree (no cross traffic).}
    \label{fig:f4}
\end{minipage}
\begin{minipage}{3.48in}
    \centering
    \includegraphics{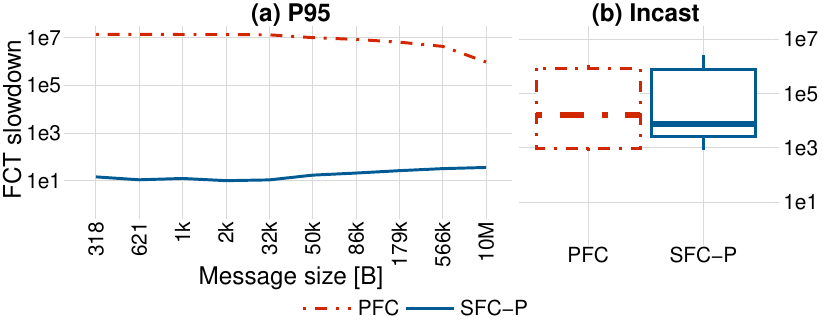}
    \captionsetup{margin={0pt,10pt}}
    \caption{Testbed: PFC causing HoLB. Workload: Hadoop 30\%; 24:1 incast.}
    \label{fig:f5}
\end{minipage}
\vspace{-0.2in}
\end{figure*}

\paragraph{Workloads and metrics.}
We adopt a similar approach as prior works~\cite{homa, hpcc, onramp, bfc} to generate all-to-all workloads based on message size distributions of well-known datacenter workloads: Google\_RPC ({\it RPC})~\cite{w3}, Facebook\_Hadoop ({\it Hadoop})~\cite{w4} and WebSearch ({\it WebSearch})~\cite{w5} for both the system evaluation as well as simulations. 
On top of these workloads, we generate an additional artificial incast load with $250KB$ messages.

We use flow completion time (FCT) slowdown (FCT slowdown) as our application performance metric.
The FCT is measured from the time the message is ready to be pushed to the networking stack to the time when the last packet of that message gets acknowledged at the sender.
The slowdown is calculated by normalizing the FCT by the assumed transfer duration at line rate in an otherwise idle network.

\subsection{System Evaluation on a 48-node Testbed}~\label{eval:testbed}
\spfc (that coverts BTS to PFC to work with today's RDMA NICs) is evaluated over $48$ host machines connected by three Tofino~2
switches~\cite{tofino2}.
The switches form a 2-ToR, 1-Spine topology. 
Each ToR connects $24$ $100G$ hosts to the spine via four $400$G inter-switch links with 3:2 over-subscription.
ECMP is used for network load balancing over the inter-switch links.
The switches are configured to use dynamic thresholding~\cite{choudhury1998dynamic}
with a setting that allows a single queue to take up to 50\%
of the shared buffer when the queue is the only one congested. 
On par with typical 3.2T switches~\cite{buffer-size}, the shared buffer
size is configured at $16$MB\footnote{Tofino2 has more buffer
but we configure it conservatively as we use only 4~Tbps
bandwidth out of its 12Tbps capacity.}.
The base RTT is $6\mu s$ within a rack and $10\mu s$ across two racks. 
We use the hosts' RDMA NIC's DCQCN as transport protocol. We compare SFC-P with PFC, the widely used flow control for RoCEv2 in practical system.  

\paragraph{SFC-P better handles large incast than standard PFC.} 
To confirm the efficacy of SFC-P we run incast $8$, $24$ and $40$ incast senders.
To increase the scale of the incast, we send multiple parallel flows per sender \cite{swift}, each sender and receiver pair uses $25$ flows (RDMA QPs) to test a large incast with a total of $200$ to $1000$ flows.

Fig.~\ref{fig:f4} plots the buffer usage of the incast receiver port measured every $1$ms. (Since there is no cross-traffic,
this queue depth is equal to the total shared buffer usage.)
As seen in Fig.~\ref{fig:f4}, \spfc yields consistently lower queuing than PFC for varying number of incast senders. 
\spfc reduces the queuing exactly down to feedback\_loop\_delay $\times$ number\_of\_hosts $\times$ 100Gbps.
The average \spfc feedback loop delay is $6\mu s$, which results in the $3MB$ stable queue depth for the 40:1 incast ratio as in Fig.~\ref{fig:f4}(c). 
Since there is only one congested queue in the switch, PFC with dynamic threshold only responds when its shared buffer usage is high, leading to higher queue depth than \spfc.

For the incast ratio of $8$:$1$, 
the buffer usage oscillates as the NIC DCQCN fails to stabilize in its interaction with flow control~\cite{pcn}. 
To effectively handle a large incast, congestion control should
be able to reduce its congestion window size to fractions of the MTU size
so that the aggregate in-flight bytes (sum of the window
sizes of all incast flows) can fit into one BDP.
HPCC and Swift~\cite{hpcc, swift} propose this approach, but the testbed's RDMA NICs do not implement it and hence cannot reach a stable state.
As more QPs and incast senders join the contention in
Fig.~\ref{fig:f4}(b) and (c), 1) the buffer usage increases
and, 2) flow control plays a major role in switch queuing, hence the effect of the coarse-grained congestion window (queue depth oscillation) does not manifest.

\paragraph{SFC-P avoids HoL blocking.}
To evaluate the impact of HoL blocking, we overlap an all-to-all workload across $16$ nodes with $8$ nodes at each switch. We then add a $24:1$ incast between switches from another 25 nodes.
The incast load is split to use $16$ senders on the remote switch and $8$ senders on the local switch and use $40$ flows per sender, for a total of $940$ flows at the receiver.
The all-to-all workload creates $4$ QPs for a sender-receiver node pair, i.e., a total of $1900$ flows in the network and a majority of them creating contention on the inter-switch links.
When standard PFC is used, PFC triggers at the incast receiver NIC as well as at the incast receiver $ToR$ switch, causing HoL blocking on the sender $ToR$-to-spine links.

Fig.~\ref{fig:f5}(a) shows the \pct{95} FCT with 30\% all-to-all traffic of the \wf workload on the 16 nodes.
\spfc significantly outperforms PFC in terms of FCT because PFC's HoL blocking slows down the entire network
(we omit other workloads as the results are similar).
As \spfc directly pauses the NIC queues at the sender, it avoids HoL blocking at inter-switch links, making the
large incast traffic have little impact on the FCTs of the all-to-all traffic. 
The FCT of the incast traffic in Fig.~\ref{fig:f5}(b) show comparable performance between \spfc and PFC. 
This is expected since the FCT of incast traffic is governed by the bottleneck link bandwidth.
There is no packet loss with PFC nor \spfc in this experiment.

\begin{figure*}[!htb]
\begin{minipage}{1.0\linewidth}
    \centering
    \includegraphics[width=1.0\textwidth]{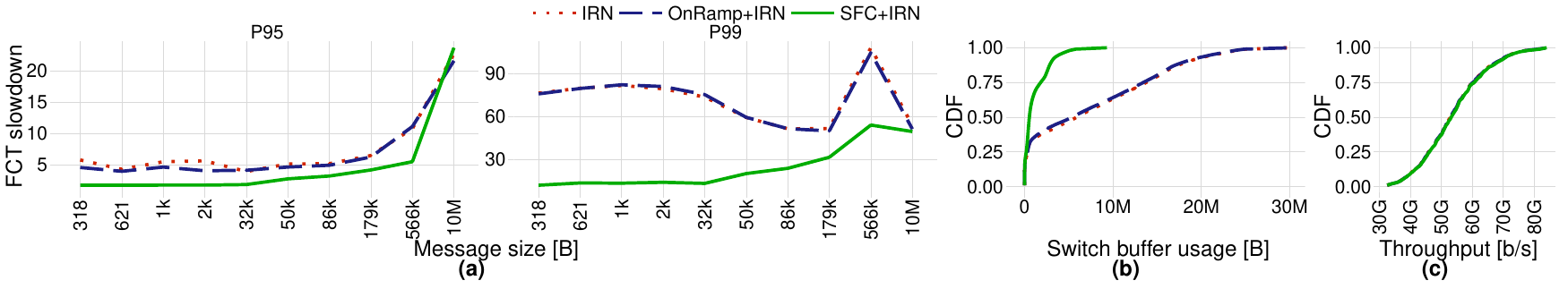} 
    \caption{{\bf Hadoop} workload with 50\% load; 8\% 128:1 incast; using {\it \bf HPCC} for congestion control.} 
    \label{fig:eval-result-center-hadoop-50}
    \end{minipage}   
    \vspace{-0.2in}
\end{figure*}
\begin{figure*}[!htb]
\begin{minipage}{1.0\linewidth}
    \centering
    \includegraphics[width=1.0\textwidth]{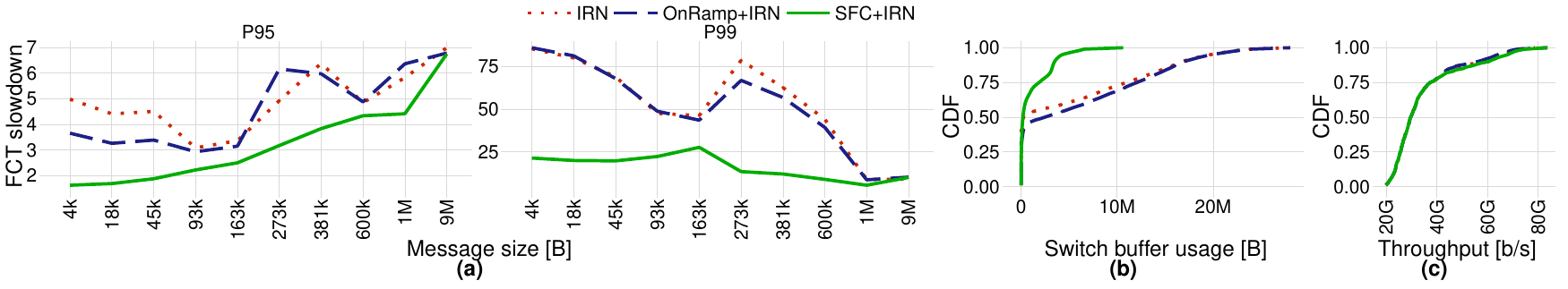} 
    \caption{{\bf RPC} workload with 30\% load; 8\% 128:1 incast; using {\it \bf HPCC} for congestion control.} 
    \label{fig:eval-result-center-rpc-30}
    \end{minipage}   
    \vspace{-0.2in}
\end{figure*}

\subsection{Scale Simulation with 512 nodes}~\label{eval:simulation}~\label{subsub:sim-perf}
We run ns3~\cite{ns3-hpcc} simulations to mainly (i) assess the
performance of \sys over the state-of-the-art flow control (\onramp), congestion control (HPCC) and loss recovery (IRN) at scale, and (ii) understand the robustness of the proposed system. A comparison with NDP is also conducted.

The topology is a 2-tier Clos of $8$ core switches, $16$ ToRs and $512$ servers similar as BFC and Homa~\cite{bfc,homa}. 
The inter-switch links are 400GbE; server-to-ToR links are $100$GbE.
ECMP is used for load balancing.
Following the link delay analysis in Poseidon~\cite{wangposeidon}, we assume $3m$ server to switch links with FEC and $500ns$ NIC processing delay yielding a total of $1\mu s$ of server-switch hop delay. 
$1.5\mu s$ switch-switch hop delay is modeled with FEC, $120m$ inter-switch-links and $600ns$ switch pipe delay.
The base RTT of a 3-hop path is $10\mu s$.
Following our testbed system, the switches implement a shared buffer with dynamic threshold~\cite{egress-shared-buffer} allowing a single queue to use up to 50\% of the $32$MB buffer before tail drop, when there is no other congested queue.

We enable \sys \emph{only at the ToR layer} to see its benefit in a practical brown-field deployment scenario.  
As network congestion is mainly located at the last hop towards the receivers under our tested workloads, this ToR-only deployment brings the most of the benefit compared to SFC deployed at every switch layer, which is verified in our simulations.

Note that real systems do not have perfectly synchronized incasts.
We experimented coordinated incasts via pull-based RDMA Read operations in a rack-level testbed of 10 servers and observed $100\mu s$ as the gap between the fastest and the slowest incast arrivals from the senders. 
In the following simulations, unless specified otherwise, we randomly start incast message transfers within an interval of $100\mu s$ which is $10\times$ the base RTT of the network.

\textbf{Default setup}. Throughout the simulation section, we use \wf at 50\% load with 128:1 incast at 8\% with $250kB$ message size and $100\mu s$ synchronization interval.
The transport protocol uses HPCC for CC and IRN for loss recovery.
When we deviate from the default setting, we highlight the changed parameters in the figure caption.

\subsubsection{Performance Evaluation}~\label{eval:simulation-perf}

\paragraph{\sys outperforms OnRamp.}
We compare \sys with OnRamp flow control and the case with no flow control while IRN and HPCC as the baseline loss recovery and CC. Three workloads are used: 50\% \wf in Fig.~\ref{fig:eval-result-center-hadoop-50}, 30\% \wg in Fig.~\ref{fig:eval-result-center-rpc-30}, and 50\% \ww in Fig.~\ref{fig:eval-result-center-websearch-50} (in appendix) each with added 8\% 128:1 incast load. 
Since the results for the latter two are similar, we depict the results of 30\% \wg here and the 50\% \ww results in the appendix.
The dotted line (IRN) in the figures is the case with no flow control. Each figure has three sub figures: (a) sub-figure plots the \emph{all-to-all} traffic's FCT slowdown for the $95^{th}$ and $99^{th}$ percentiles (P$95$ and P$99$) by message sizes, which are grouped into 10 bins.
Sub-figure (b) is the CDF of the per-switch shared buffer usage (watermark sampling every $10\mu s$) and (c) depicts the CDF of all receivers' throughput.

We observe the following common patterns across the three workloads: 
(1) SFC consumes significantly less switch buffer than IRN (no flow control) or OnRamp + IRN.
(2) SFC nor OnRamp change the throughput distributions, meaning they do not cause switch under-run.
(3) Lower buffering and the same high throughput of SFC leads to significantly improved FCT. 
Overall, the gain is larger at P99 (2x-6x) than P95 (1.2x-2x). 
This confirms the benefit of lower buffering on the `tail' latency of cross traffic. 
(4) The FCT gain is larger at small to median size messages as their FCT is dominated by queueing delay while for large messages FCT is a function of throughput.

Note that the gain of \sys over standard RoCEv2 mechanisms -- PFC and Go-back-N loss recovery -- are also evaluated, often resulting in 2 orders of magnitude difference. We omit them in the graphs to focus on the improvement that SFC makes atop the state-of-the-art: OnRamp and IRN.

\paragraph{SFC significantly reduces the switch buffer usage.}
The peak buffer usage of OnRamp and IRN (no FC) reaches slightly below the $32MB$ mark,
meaning packet drops in this typical switch buffer model of $32MB$. Dynamic threshold doesn't allow any one or small number of queues
to consume the entire shared buffer. Depending on the concurrent number of congested queues, drops are possible anywhere in the
range of $16MB$ to $32MB$ of buffer usage. \sys contains the peak buffer usage $2.5-3\times$ lower than the others, well below 16MB. 
It demonstrates that lossless networking, while not guaranteed for all workloads, with great FCT performance is possible without PFC for representative DC workloads with heavy incasts.

\paragraph{SFC helps with microbursts (no incast) or smaller incast.}~\label{eval:other-workloads}
\begin{figure*}[!htb]
\begin{minipage}[t]{2.38in}
    \centering    
    \includegraphics[width=1\linewidth]{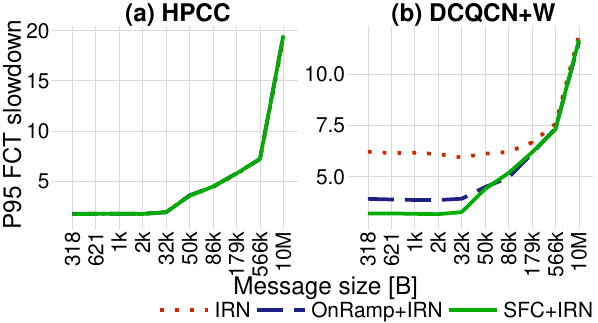}
    \caption{\small {\bf Hadoop} at \underline{80\% load}; \underline{no incast}.}
    \label{fig:eval-detail-hadoop-80-il0}
\end{minipage}      
\begin{minipage}[t]{4.58in}
    \centering    
    \includegraphics[width=1\linewidth]{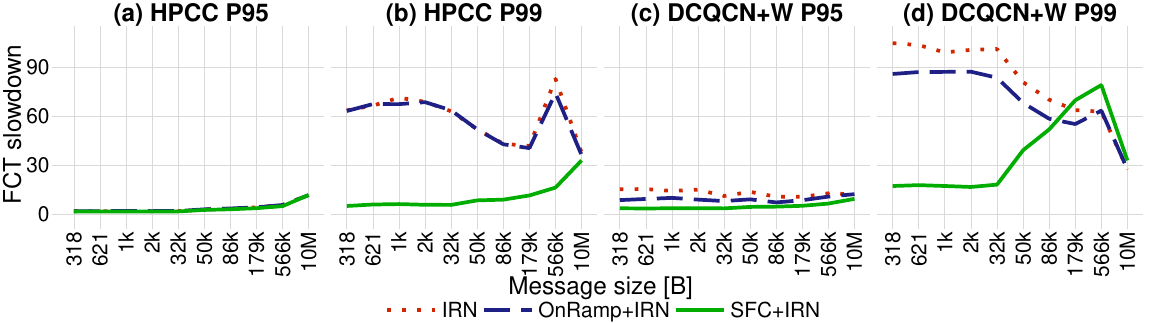}
    \caption{\small {\bf Hadoop} at 50\% load; \underline{5\%} 128:1 incast.}
    \label{fig:eval-detail-hadoop-50-il5}
\end{minipage} 
\end{figure*}
Though we have explained the benefit of SFC in the context of heavy incast, SFC helps microbursts naturally created by all-to-all traffic. 
Fig.~\ref{fig:eval-detail-hadoop-80-il0} depicts the \pct{95} FCT slowdown for a 80\% Hadoop workload `without' incast, while using HPCC (a) and DCQCN+W (b) as congestion control algorithms.
HPCC handles the 80\% load without needing flow control as depicted in Fig.~\ref{fig:eval-detail-hadoop-80-il0}(a): all mechanisms perform similar with no significant differences throughout all message sizes.
Compared to HPCC, DQCCN+W reacts to microbursts slower and it fails to keep the equilibrium buffer as low as HPCC, leaving a room for flow control to help. OnRamp reduces FCT slowdown by up to 60\% compared to IRN (no FC) and \sys further improves FCT by 20\% for message sizes below 32KB.

In Fig.~\ref{fig:eval-detail-hadoop-50-il5}, we roll back to 50\% load of Hadoop and use 
5\% incast, a more moderate than the 8\% incast used in Fig.~\ref{fig:eval-result-center-hadoop-50} to Fig.~\ref{fig:eval-result-center-websearch-50}. HPCC and DCQCN+W are used as CC.
We use the same y-axis scale for all sub figures to fairly compare the tail FCT slowdowns.
With HPCC, the \pct{95} FCT slowdown of Fig.~\ref{fig:eval-detail-hadoop-50-il5}(a) is marginally improved by \sys compared to the alternatives. This shows the efficacy of INT-based CC in handling incast, compared to DCQCN+W \pct{95}.
Yet in Fig.~\ref{fig:eval-detail-hadoop-50-il5}(b), SFC shows a significant improvement of \pct{99} FCT slowdown compared to the alternatives using HPCC.
Note that OnRamp meaningfully improves the tail FCT only for DCQCN+W, not much for HPCC across different workloads (Fig.~\ref{fig:eval-result-center-hadoop-50} - Fig.~\ref{fig:eval-detail-hadoop-50-il5}), which echoes the OnRamp paper's conclusion: 
\enquote{the performance of HPCC is not significantly improved because it utilizes recent and detailed congestion information from the network elements and is already highly performant.}~\cite{onramp}.

\subsubsection{Robustness Evaluation.}~\label{eval:simulation-robust}
\begin{figure*}[!htb]
\begin{minipage}[t]{1.48in}
    \centering    
    \includegraphics[width=1\linewidth]{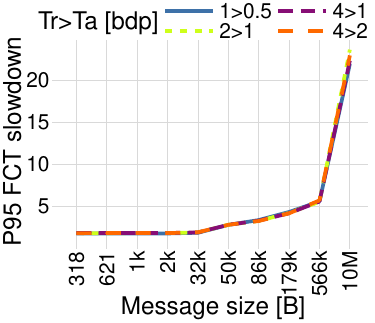}
    \captionsetup{margin={0pt,3pt}}
    \caption{\small Impact of \sys settings. Default setup.}
    \label{fig:vary_parameter}
\end{minipage}    
\begin{minipage}[t]{2.48in}
    \centering
    \includegraphics[width=1\linewidth]{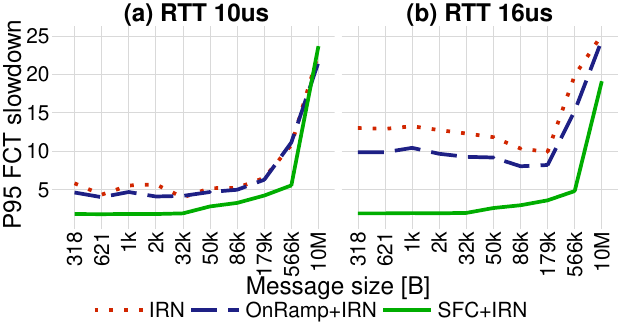}
    \captionsetup{margin={3pt,3pt}}
    \caption{\small Impact of longer RTT. Default setup.}
    \label{fig:vary_rtt}
\end{minipage} 
\begin{minipage}[t]{2.98in}
    \centering
    \includegraphics[width=1\linewidth]{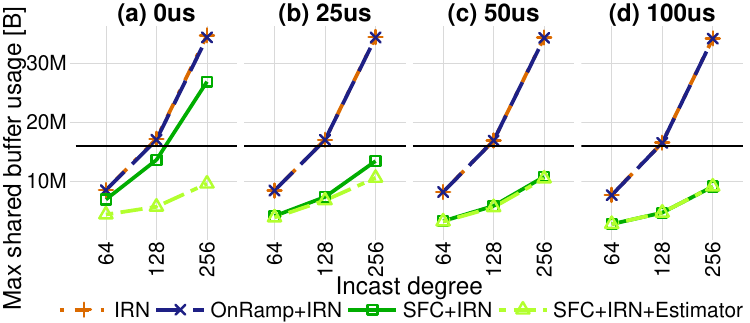} 
    \captionsetup{margin={3pt,0pt}}
    \caption{\small Incast-only workload with varying incast degree, and incast synchronization interval. HPCC.}
    \label{fig:vary-incast-sync}
\end{minipage}  
\vspace{-0.2in}
\end{figure*}

We evaluate the robustness of \sys by studying the effect of
various system and simulation parameter changes on its performance.

\paragraph{\sys is insensitive to parameter tuning.}
Parameter settings can significantly impede the performance of CC algorithms~\cite{hpcc} as well as flow control algorithms~\cite{pfc, irn}.
We investigate the behavior of \sys when using parameters from the range
of values that are expected to be practical. 
\sys uses two parameters: (1) the queue depth threshold to trigger
\sys; (2) the target queue depth that \sys drains down to. 
We analyzed these parameters and their values in \S\ref{subsec:egress} and concluded that the recommended parameters values are $2\times$ BDP for the trigger threshold and $1\times$ BDP for the target queue depth.
As we deploy \sys only at ToR switches, the feedback loop delay is approximated by the network-wide RTT with a resulting BDP of $125$KB. 

We test the robustness over the target queue depths of $0.5$~BDP, one BDP and $2$~BDP bytes, and set the trigger thresholds as $2$x of the target queue depth. 
We take the upper bound of the trigger threshold of $4$ BDP, as the aggregation of all the queues of the $4$ BDP in a switch are close to full switch buffer. 
The four pairs of trigger threshold and target queue depth are tested in
in Fig.~\ref{fig:vary_parameter} and show no significant differences in the \pct{95} FCT slowdown,
showing the robustness of \sys.

\paragraph{\sys performs consistently in larger RTT networks.}
Ingress BTS alone is expected to get less effective with increasing feedback delay, as
more traffic is injected into the network before BTS can slow it down.
\sys uses near-source caching to help in this case. 
Fig.~\ref{fig:vary_rtt} shows the \pct{95} FCT slowdown for the \wf
workload as the RTT increases from $10\mu s$ (a) to $16\mu s$ (b). 
The FCT slowdown of \sys remains virtually unchanged as RTT increases while the other schemes' (\onramp and no FC) FCT increases significantly as a function of RTT.
This is mainly because \sys's pause signals are not only have a lower latency than the alternatives, but caching effectively offsets the impact of the higher RTT.

\paragraph{Impact of various incast synchronization and degree.}
The efficacy of the cache depends on the synchronization of the incast, i.e., the time window during which the incast flows start sending.
In Fig.~\ref{fig:vary-incast-sync}, we vary the incast synchronization and the incast degree to see their impact on the maximum shared buffer usage of any switch in the network. We assume infinite buffer 
to understand the switch buffer utilization performance without the distorting impact of tail drops or affected by RTO setting.
To get a sense on the possible congestion drops, black horizontal lines are added to indicate the $16MB$ queue depth where drop would start to happen if we didn't assume infinite buffer. Only the incast victim queue has traffic in this experiment and it could grow up to the half of the 32MB shared buffer by dynamic threshold. 
The incast degree varies across $64$, $128$, $256$.
The incast synchronization is varied from the theoretic worst case of perfect synchronization ($0us$ synchronization) in Fig.~\ref{fig:vary-incast-sync}(a) to $100us$ in Fig.~\ref{fig:vary-incast-sync} (d), which represents $10\times$ base RTT.

Fig.~\ref{fig:vary-incast-sync} clearly shows that SFC better manages the switch buffer under various incast patterns: 1) smaller incast degree of 64:1 or 2) extremely large (256:1 out of 512-node cluster) and perfectly synchronized. The benefit of SFC is larger as the synchronization interval increases. In the case of perfectly synchronized incasts, SFC caching doesn't help as much and the near-source local estimator (the optional feature presented in Section~\ref{subsec:egress}) plays a key role in bounding the max buffer usage under the 16MB line. As expected, the local estimator is not needed if the incast starts are synchronized to 25us (2.5x base RTT) or higher.  

\textbf{Appendix~\S\ref{app:eval}} provides additional evaluations supporting the followings: (1) SFC consumes minimal amount of switch resources, (2) SFC does not change the fairness property of underlying CC, (3) SFC with BTS better handles extreme incasts than NDP with trimming + RTS (Return-To-Sender).

%% file: sections/appendix.tex
\newpage
\section*{Appendix}


\section{Discussion \& Future Work}
\label{app:discuss}

\paragraph{Lossless vs. Lossy.} There is an on-going debate in industry
around the need for lossless fabrics, as opposed to lossy ones. In our
opinion, lossless fabrics (with respect to congestion loss) became
necessary because end-to-end transport failed to avoid congestion drops
(e.g., use of imprecise congestion signals like drop, dupACK, smoothed
RTT) or to efficiently recover from drops (e.g., RDMA go-back-N).
While SFC is \emph{not} intended as a mechanism to achieve a lossless fabric,
BTS does allow the NIC/host stack to minimize or avoid congestion drops and to
handle anomalous events more quickly.

\paragraph{Reacting to NIC congestion.}
\label{subsec:nic-congestion}
NIC congestion occurs for various reasons: from incast to
high packet-per-second bursts, and PCIe and memory bandwidth
bottlenecks~\cite{swift, jiang2020, collie}. 
When used with PFC, the NIC congestion triggers PFC backpressure to ToR switches that may have multiple
concurrent uplink congestion points due to ECMP collisions of
constant high-load flows leading to performance degradation. 

SFC reacts to NIC congestion signaled by PFC.
PFC frames sent from the NIC pause the transmission of packets by the switch egress queue, which automatically translates to queue buildup that triggers BTS, reducing congestion and PFC spreading.
Triggering BTS upon receiving NIC-generated PFC (even before queue buildup) is a potential optimization for \sys.

\paragraph{Number of competing senders.}
The bloom filter used in BTS suppression can be easily extended to 
a counting bloom filter and track the number of competing incast flows or senders.
One can use the information to set the pause time more aggressively and further reduce the peak queue depth.
We leave it to a future work.

\section{Evaluation}
\label{app:eval}

\subsection{System evaluation details}
\label{eval:system-hw-details}
The Tofino 2 switches are running SONiC 202201 (SONiC.HEAD.0-dirty-20220127.163606). 
Each host has an Intel Xeon E5-2697A v4 @ 2.60GHz, 64GB RAM running Redhat 8.4 
(kernel v$4.18.0$), and an Intel E810 100G RDMA 
NIC~\cite{intele810} running driver v$1.8.2$.  

\begin{figure*}[!htb]
\begin{minipage}{1.0\linewidth}
    \centering
    \includegraphics[width=1.0\textwidth]{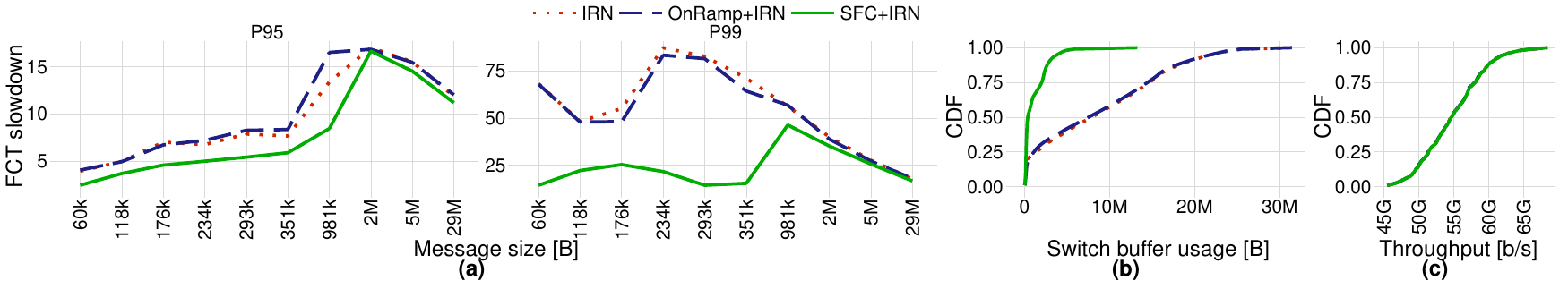} 
    \caption{{\bf WebSearch} workload with 50\% load; 8\% 128-to-1 incast; using {\it \bf HPCC} for congestion control. } 
    \label{fig:eval-result-center-websearch-50}
    \end{minipage}   
    \vspace{-0.2in}
\end{figure*}

\subsection{\sys minimally consumes switch resources}~\label{eval:simulation-overhead}

\begin{table}[!h]
\centering
\small
\begin{tabular}{lrrl}
  \hline
Setup & Suppr. & Cache & BTS reduction \\ 
  \hline
  RPC  & 100 &  12 & 77.48\% \\
  Hadoop & 105 &  12 & 81.91\% \\ 

   \hline
\end{tabular}
\caption{RPC (Fig.~\ref{fig:eval-result-center-rpc-30}) and Hadoop (Fig. ~\ref{fig:eval-result-center-hadoop-50})}\label{tab:tab2_1}
\end{table}

\noindent \sys costs switch memory for \sqp suppression,
SFC Pause cache.
As switch memory is limited, too much memory consumption can limit the scale of \sys deployment
and hurt \sys's performance.
Thus, we monitor the two tables
for their maximum entry occupancy and collect the BTS reduction ratio over all the \sys-enabled switches
in the same large-scale experiments in Figure~\ref{fig:eval-result-center-rpc-30} and Figure~\ref{fig:eval-result-center-hadoop-50}.
The corresponding results are shown in  Table~\ref{tab:tab2_1}.
The max occupancy of the BTS suppression bloom filter is small, i.e., less than $110$ entries in any of the scenarios. 
For the minimal 3-hash filter design we use, the false positive is near-zero. 
In addition, with the bloom filter, \sys reduced the \sqps between $77-81\%$.
The Pause cache table has less than $15$ entries in all configurations; its memory consumption is minor.

Note that these stateful tables have mechanisms to immediately detect unnecessary entries and retire them:
frequent reset, pause-end time.
Moreover, the two tables are for optimizations. If the table is full, or a hash collision
happens, we just don't apply the optimization to newly arriving BTS packets.

\subsection{SFC does not impact flow fairness}~\label{eval:cc-fairness}
SFC pauses every flow sharing the same congestion point fairly in time. Thick elephant flows may get paused more in terms of BPS rate than mice flows, but all of them get the same pause time duration by BTS. SFC simply migrates the queuing location from the switch buffer to each sender's buffer without changing the minimally required buffering time, hence there is no negative effect on the fairness property that underline CC provides for long flows.

\begin{figure}[!htb]
    \centering    
    \includegraphics{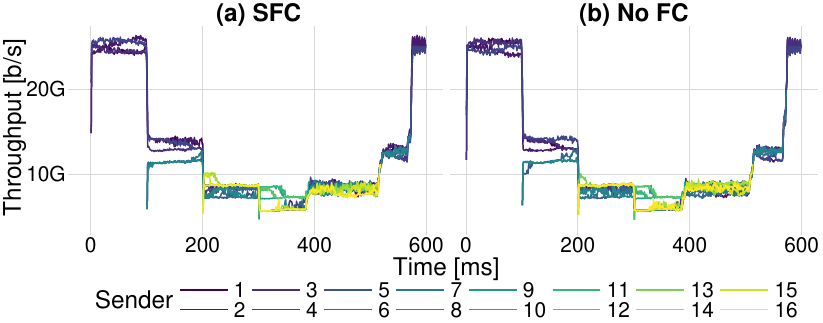}
    \caption{\small SFC has zero impact on fairness. DCQCN+W.}
    \label{fig:fairness}
\end{figure}

Fig.~\ref{fig:fairness} demonstrates it by comparing the bandwidth sharing and throughput convergence of competing flows, (a) with SFC and (b) without it. Four groups of senders, each with four senders, joins and leaves a contention on 100G link one by one. DCQCN+W is used as CC and SFC gets triggered when a new group joins. 
As expected and desired, SFC has zero impact on the fairness.
We also tried HPCC and SFC was never triggered in this mild congestion scenario.

\subsection{Comparison with NDP}~\label{eval:ndp}
As discussed in \S\ref{sec:backward-feedback}, NDP trimming is a solution to handle high incast, well suited for receiver-driven transports. Meanwhile, many existing and widely-deployed transports heavily run sender-driven congestion control and scheduling, which we target in this paper. 
Sender-driven schedulers handle traffic mix and heterogeneous
topology well. A question is \emph{how best can a sender-driven approach handle heavy incast and how well do that compare with a receiver-driven approach?} Here, we compare NDP and SFC 
and show that BTS/SFC leverages existing switches' shared buffers (compared to NDP's aggressive queue sizing) and avoids 
incast drops by keeping the buffer consumption low
enough to fit in shallow buffers of commodity DC
switches.

\begin{figure}[!htb]
    \centering
    \includegraphics{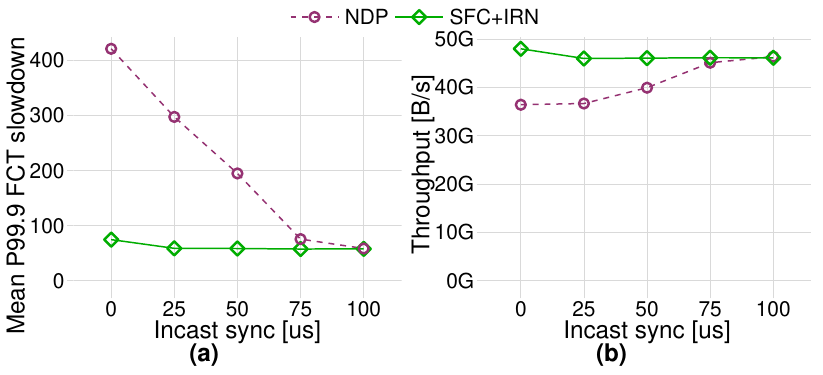}
    \caption{{\bf Hadoop} workload with 50\% load; 8\% 128:1 incast; using {\it \bf HPCC} for congestion control.} 
    \label{fig:ndp}
\end{figure}

Fig.~\ref{fig:ndp} compares NDP and SFC in mean throughput and tail FCT slowdown metrics. Working with NDP authors, we improved the htsim implementation of NDP~\cite{ndp-code} by adding 1) byte level credits (vs packet level) to efficiently schedule small messages below MTU, 2) flow-level round robin at the sender to avoid HoLB, 3) read the same flow trace inputs as the ns3.
The same default simulation setup and workload traces (50\% Hadoop and 8\% 128:1 incast) from \S\ref{eval:simulation} are used for both SFC and NDP, while varying the incast synchronization interval. Although we made many efforts/improvements to fairly compare NDP results from htsim~\cite{ndp} and SFC results from NS3, we avoid comparing their results in fine-grained details. (Note Homa~\cite{homa} also plotted results from two different simulators -- htsim for NDP and OMNeT++ for other schemes -- in one graph.)

We make two high-level observations from Fig.~\ref{fig:ndp}: (1) NDP and SFC perform quite closely each other when the incast synchronization is not too tight, especially on P95 and P99 FCT metrics (not shown as they are similar between SFC and NDP). NDP scheduler efficiently re-schedules initial incast drops and achieve high throughput and low FCT at part with SFC. (2) When incast senders are tightly synchronized, the high-priority control packet queues of NDP switches -- that serve trimmed headers, ACKs, NACKs and credit packets -- are overflowed. NDP sets the size of control packet queue at 1x BDP. The loss of credit packets lead to throughput degradation (Fig.~\ref{fig:ndp}(b)) and timeouts of small single-packet messages, hurting their P99.9 tail latency (Fig.~\ref{fig:ndp}(a)). The high slowdown is inflated by the default 50ms timeout parameter used by the NDP simulator. 

Tuning the control queue size and timeout parameter should help but the fundamental of dropping control packets upon high incast is there with the aggressive queue size selection of NDP. NDP switches generate RTS (Return-to-Sender) when trimmed headers are dropped but RTS doesn't help with the loss of other control packets. 

Note the LPF-based local estimator is disabled in SFC switches as SFC with caching alone manages the shared buffer usage under the drop point even with the perfectly synchronized incast of 128 senders, as observed in Fig.~\ref{fig:vary-incast-sync}.

\input{sections/appendix_theory}

\null
\vfill
\vspace{20em}

%% file: sections/appendix_theory.tex
\section{Theoretical Analysis} \label{app:theory}

We hypothesize that a reduced signaling delay such as the sub-RTT feedback primarily helps shrinking buffers in the network as it would allow senders to detect and react to congestion much earlier, i.e., before a large queue builds up.
Therefore, we ask \textit{What is the minimum buffer size a switch should have given a feedback delay such that it can tolerate a congestion event without dropping any packets?} 
We will assume optimal flow control and congestion control algorithms which set the transmission rate/cwnd to the fair share immediately after receiving the first Congestion Notification (CN) after pausing an ideal amount of time to drain the existing congestion.

\begin{figure}[htbp]
\centering
\includegraphics[width=0.7\columnwidth]{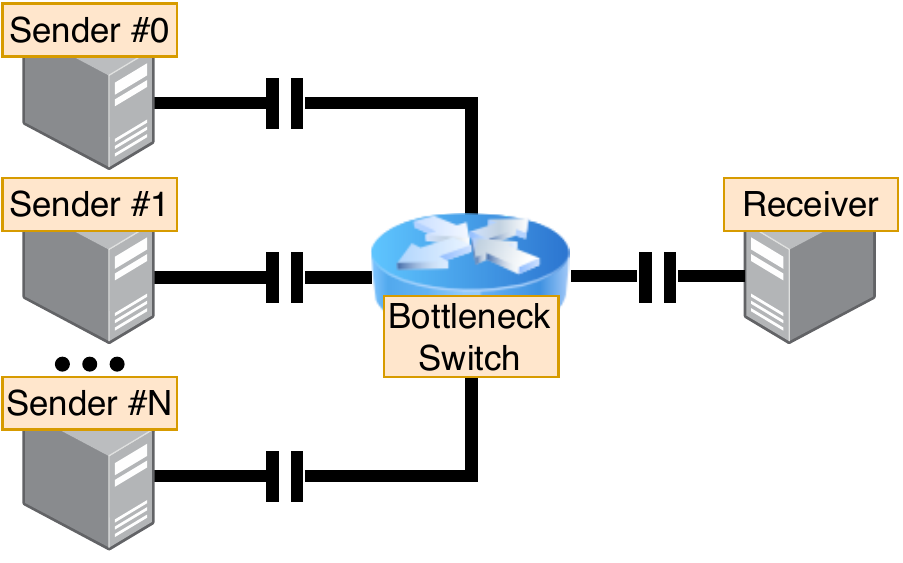}
\caption{Simplified topology for the theoretical analysis.}
\label{app:fig:dumbbell}
\end{figure}

Imagine the topology in Figure~\ref{app:fig:dumbbell} where there are N+1 flows running through the same bottleneck switch towards the receiver. 
Note that this topology is not necessarily a two hop star topology. 
Plus, the bottleneck switch does not need to be the last hop for this analysis. 
The links shown on the figure abstract away all the other networking elements and models them as a single link since they do not mandate flow’s transmission rate/cwnd.

We define the following:
\begin{itemize}
    \item \textbf{$R$}: Bottleneck link capacity (i.e., bits per second)
    \item \textbf{$T_A$}: Propagation delay between the senders and the bottleneck switch. Every flow may have a different delay to the bottleneck switch, but we consider the largest among senders for the worst case analysis.
    \item \textbf{$T_B$}: Propagation delay between the bottleneck switch and the receiver.
    \item \textbf{$T_{ToR}$}: The propagation delay between the sender and the immediate ToR switch connected to the sender.
    \item \textbf{$T_S$}: The serialization delay of an MTU size packet. We assume control packets have zero serialization delay.
    \item \textbf{$RTT$} $= 2T_A + 2T_B + 2T_S$. 
    \item \textbf{$RTT_{BTS}$} $= 2T_A + T_S$ =\textbf{$T_F$} .
    \item \textbf{$k$}: The target number of packets in a buffer before a CN (Congestion Notification) is triggered (\ie ${Q_T}_h/MTU$).
\end{itemize}

We start with a simple scenario and progressively generalize the case.

\subsection{2 Flow Congestion} \label{app:theory:2-flow}

Suppose Sender\#0 is at steady state, transmitting at the line rate without queuing at the bottleneck. 
At $t=0$, Sender\#1 starts at line rate.

The first CN will be generated when the $k^{th}$ packet of Sender\#1 arrives at the switch. 
Without BTS, the CN is reflected by the receiver with ACK packets after waiting in the queue first, \eg OWD for OnRamp~\cite{onramp}.
Therefore the first CN will be observed by a sender at $t=2T_A + 2k\times T_S + 2T_B$.
Even if the senders immediately decrease their rate/cwnd to the new fair share, all the packets Sender\#1 has sent so far are going to create queuing at the bottleneck switch.
The size of the buffer required to accommodate all these packets, $B_{ack}$, can be calculated as:

\begin{equation} \label{eq:b-ack}
    B_{ack} = R \times (2T_A + 2k\times T_S + 2T_B)
\end{equation}

In the case of BTS, the sender will receive the first CN at $t = 2T_A + k\times T_S$.
Therefore, the required amount of buffer for accommodating the congestion becomes

\begin{equation} \label{eq:b-bts}
    B_{bts} = R \times (2T_A + k\times T_S) 
\end{equation}

Note that $B_{bts} < B_{ack}$ and the saving for buffer space is $R\times (2T_B + k\times T_S)$.
For a scenario where $R = 100Gbps$, $RTT= 10\mu$s, $T_B = 1\mu s$, and $MTU = 4000$B, this saving is $129$KB or $10.32\mu$s which corresponds to 38.7\% reduction in the maximum congestion.\footnote{$T_A = 4\mu$s and take $k={Q_T}_g/MTU=T_F\times R / MTU \approx 26$}
\subsection{N Flow Incast} \label{app:theory:n-flow-incast}

This time, suppose $N>k$ additional flows start at line rate at $t=0$ while Sender\#0 is at steady state.
Therefore the first CN would be emitted by the switch when the first packets of each sender arrive at the bottleneck switch which will be observed by the senders at $t=2T_A + T_S = T_F$.
Then,

\begin{equation} \label{eq:b-incast}
    B_{incast} = N \times R \times (2T_A + T_S) = N\times R\times T_F
\end{equation}

Without BTS, the first CN would be delivered to one of the senders at $t= RTT$ whereas other senders would receive a CN later depending on the order in which their data packets arrive at the congested queue which implies that $B_{ack} > N\times R\times RTT$.
Therefore, the buffer space required without BTS would be more than $RTT / T_F$ times higher compared to the use of BTS packets.

Due to the nature of how distributed systems work, incasts in the wild are always asynchronous. While evaluating \sys, we observed that the inter-arrival time of incast flows is a multiple ($m>1$) of the $RTT$.
When 
\begin{equation} \label{eq:asynch-incast-ineq}
    m\times RTT > T_F + T_P
\end{equation}
ideal flows would receive CN ($T_F$), pause enough to drain the queue ($T_P)$, and adjust their rate/cwnd to the fair share before new flows join the network.
Therefore, the buffer utilization of an asynchronous incast would be equivalent to the scenario described in \S\ref{app:theory:2-flow}.

Note that 
\begin{equation} \label{eq:asynch-incast-inter-arrival}
    m\times RTT = m\times B_{ack} / R
\end{equation}
\begin{equation} \label{eq:asynch-incast-queue-drain-time}
    T_F + T_P = T_F + (B_{bts} - {Q_T}_g) / R = B_{bts}/R
\end{equation}
Inserting these equalities into inequality~\ref{eq:asynch-incast-ineq} gives us the conclusion that the queue utilization would be limited to $B_{bts}$ as long as $m > B_{bts} / B_{ack}$ or the inter arrival time of flows ($m\times RTT$) is longer than $B_{bts} / R$. 
For the numerical example given in \S\ref{app:theory:2-flow}, $B_{bts} / R$ would be calculated as $16.32\mu$s.

\subsection{The Effect of Congestion Caching} \label{app:theory:congestion-caching}

Suppose $N>k$ flows start at line rate with a Poisson inter-arrival time of $\lambda \geq m\times RTT \geq 2RTT$ while Sender\#0 is at steady state. 
For the sake of simplicity, let's assume $k=1$ which is the threshold value for minimum queuing.

The congestion created by Sender\#1 would be cache miss at the immediate ToR switch and the congestion feedback would fall back to the BTS generated by the bottleneck switch.
Then, the buffer requirement for this scenario would be $B_{bts}$ as calculated in \S\ref{app:theory:2-flow}.
In this case, the time when the cache entry is added onto the immediate ToR switch would be the following:

\begin{equation} \label{eq:t-cache}
    T_{cache} = T_F - T_{ToR}
\end{equation}

In order for Sender\#2 to get a cache miss under the same ToR switch, the first packet of the sender should arrive at the ToR switch before $T_{cache}$ which means it should start before $T_{cache~miss} = T_{cache} - (T_{ToR} + T_S) = 2(T_A - T_{ToR})$

If Sender\#2 starts at $T_{Sender\#2} < T_{cache~miss}$, it will therefore get the cache miss for its first packet.
However, eventually a packet from its first cwnd will get the cache hit and the sender will receive congestion feedback by the time Sender\#1 receives its first BTS, \ie $T_F$, because Sender\#1's BTS will be earlier than any BTS generated for Sender\#2. 
Therefore, Sender\#2 will emit $B_{Sender\#2} = R\times (T_F - T_{Sender\#2})$ bytes before pausing.

If $T_{Sender\#2} > T_{cache~miss}$, Sender\#2 will see a cache hit with its first packet given that the congestion by the previous sender has not already been drained yet.
In this case, the feedback delay will be $2T_{ToR} + T_S$ instead of $2T_A + T_S$ which implies that
\begin{equation} \label{eq:b-cache-hit}
    B_{cache~hit} = R\times (2T_{ToR} + T_S) < B_{bts}
\end{equation}
thanks to the caching feature of \sys which effectively reduces the feedback latency for a flow.
If there was no caching, the required buffer space would always be $T_A / T_{ToR}$ times higher than the cache hit scenario assuming a negligible $T_S$.